\documentclass[journal]{IEEEtran}

\usepackage{amsmath}
\usepackage{amsfonts}
\usepackage{mathrsfs}
\usepackage{amssymb}
\usepackage{mathtools}
\usepackage{bm}
\usepackage{cite}

\usepackage[caption=false,font=footnotesize]{subfig}
\usepackage{graphicx}
\usepackage{stfloats}
\usepackage{acronym}
\usepackage[colorlinks=true,citecolor=red,bookmarksdepth=2]{hyperref}

\newtheorem{lemma}{Lemma}

\newtheorem{theorem}{Theorem}

\allowdisplaybreaks
\usepackage{xcolor}

\begin{document}
\title{Heterogeneous Ultra-Dense Networks with Traffic Hotspots: A Unified Handover Analysis}
\author{He Zhou, Haibo Zhou, \IEEEmembership{Senior Member, IEEE}, Jianguo Li, Kai Yang,~\IEEEmembership{Member,~IEEE}, \\Jianping An,~\IEEEmembership{Member,~IEEE}, and Xuemin (Sherman) Shen, \IEEEmembership{Fellow, IEEE}
\thanks{He Zhou, Jianguo Li, and Kai Yang are with the School of Information and Electronics, Beijing Institute of Technology, Beijing, China (e-mail:willzhou@bit.edu.cn, jianguoli@bit.edu.cn, yangkai@ieee.org).

Jianping An is with School of cyberspace science and technology, Beijing Institute of Technology, Beijing, China (e-mail:an@bit.edu.cn).

Haibo Zhou is with the School of Electronic Science and Engineering, Nanjing University, Nanjing 210023, China (e-mail: haibozhou@nju.edu.cn).

Xuemin (Sherman) Shen is with the Department of Electrical and Computer Engineering, University of Waterloo, Waterloo, ON N2L 3G1, Canada (e-mail: sshen@uwaterloo.ca).}}

\maketitle

\begin{abstract}
With the ever-growing communication demands and the unceasing miniaturization of mobile devices, the Internet of Things is expanding the amount of mobile terminals to an enormous level. To deal with such numbers of communication data, plenty of base stations (BSs) need to be deployed.
However, denser deployments of heterogeneous networks (HetNets) lead to more frequent handovers, which could increase network burden and degrade the users experience, especially in traffic hotspot areas. 
In this paper, we develop a unified framework to investigate the handover performance of wireless networks with traffic hotspots. 
Using the stochastic geometry, we derive the theoretical expressions of average distances and handover metrics in HetNets, where the correlations between users and BSs in hotspots are captured. 
Specifically, the distributions of macro cells are modeled as independent Poisson point processes (PPPs), and the two tiers of small cells outside and inside the hotspots are modeled as PPP and Poisson cluster process (PCP) separately. 
A modified random waypoint (MRWP) model is also proposed to eliminate the density wave phenomenon in traditional models and to increase the accuracy of handover decision. 
By combining the PCP and MRWP model, the distributions of distances from a typical terminal to the BSs in different tiers are derived. 
Afterwards, we derive the expressions of average distances from a typical terminal to different BSs, and reveal that the handover rate, handover failure rate, and ping-pong rate are deduced as the functions of BS density, scattering variance of clustered small cell, user velocity, and threshold of triggered time. 
Simulation results verify the accuracy of the proposed analytical model and closed-form theoretical expressions.

\end{abstract}


\section{Introduction}\label{sec:introduction}

Driven by the development of Internet of Things (IoT), deploying the miniaturized terminals and dense base stations~(BSs) in terrestrial communication networks is becoming the effective solution to meet the ever-growing demands of communication in hotspot areas \cite{Arshad2016,Yang2018,Damnjanovic2011,Ni2021,Gao2022}.
In recent years, using stochastic geometry to model the spatial deployment of BSs in the IoT network has received significant attention from academics \cite{Panigrahy2018,Ge2020}.
In order to simulate the randomness of BSs distribution, the locations of BSs in different tiers are usually modeled as independent Poisson point processes~(PPPs). 
Performing the signal-to-interference-plus-noise ratio (SINR) distribution analysis of multi-tier networks at a typical user, the coverage probability and outage probability are investigated in \cite{Andrews2016}.
Although using PPP to model the random networks is tractable and convenient, it does not capture the coupling between BSs locations and user equipments (UEs) in hotspots, where the users and small BSs (SBSs) tend to be clustered. 
Since lots of achievements which focus on the PPP and further studies have been exhibited, we mainly pay our attention on the handover analysis based on the Poisson cluster processes (PCP) where few studies are proposed in this area.

In this context, we consider using PCP, which has gained much attention on communication performance analysis in traffic hotspots to model the locations of BSs \cite{Saha2019,Afshang2018a}. 
Although the PCP-based model is suitable for characterizing ultra-dense networks, the innumerable diminutive mobile devices in the IoT network will bring complex handover management problems \cite{LopezPerez2012,Bi2019,Stamou2019,Yuan2020}.
As for the handover performance under PCP-based model, the authors in \cite{Bao1,Bao2} derived the analytical expressions for the rates of horizontal and vertical handoffs experienced by an active user with arbitrary movement trajectory. 
Specifically, the authors in \cite{Bao1} considered using two tiers of PPP with different deployment intensities to model the locations of cluster centers and the cluster members. 
In the mean time, the set of $\bigtriangleup d$-extended is first introduced to quantify the one-dimensional measures of different length intensities on the two dimensional plane.
In \cite{Bao2}, a stochastic geometric analysis framework on user mobility was proposed. 
The authors also provided guidelines for optimal tier selection under various user velocities, taking both the handoff rates and the data rate into consideration.
However, the sojourn time for users which stay in the handover boundaries is not taken into consideration. 
To the authors' knowledge, only few of the prior works focus on the handover analysis under PCP and take the sojourn time into consideration.

Recent developments in the field of stochastic geometry have led to a renewed interest with handover management in wireless networks \cite{Zhang2019,Teng2018,Lee2014,Arshad2019}.
In stochastic geometry, the PPP is widely used in modeling and statistic analysis because of its universality and uniform distribution. 
In \cite{Arshad2016b}, the authors proposed a cooperative handover management scheme to reduce unnecessary handovers in dense cellular networks. 
Furthermore, a velocity-aware mathematical model and an escalated handover scheme for the users were proposed in \cite{Arshad2017}, in which the BSs along with the moving trajectory will be selected instead of those BSs which have stronger transmit power to maintain longer connection durations and to reduce handover rate.
To quantify the performance of the proposed handover schemes via stochastic geometry, the effects of mobility on the densification gain were discussed in a two-tier downlink cellular network with ultra-dense small cells and $C$-plane/$U$-plane split architecture \cite{Ibrahim2016}.
A joint bandwidth allocation and a call admission control strategy were proposed in \cite{Fang2015} to reduce unnecessary handovers.
The authors proposed joint optimization of cell association and power control that maximizes the long-term system-wide utility in \cite{Qian2017-1} to reduce the handover rate. 
A joint optimization of cell association and power control solution was also proposed in \cite{Qian2017-2} to improve the spectrum and energy efficiency and reduce handover rate.
For handover analysis, the closed-form theoretical expressions of handover rate, handover failure rate, and ping-pong rate were investigated in \cite{Xu2017}, where the distribution of BSs' locations follow independent PPP and the expressions were resolved as function of BS density and triggered time threshold. 
With redefined definitions, the handover rate in our paper decouples with the handover failure rate. 
Compared with \cite{Xu2017}, the handover rate is more specific and accurate since it only contains the influence of user mobility model and sojourn time inside the ERB circles. 
Moreover, according to the description in \cite{access2010further}, the handover failure rate in our paper is denoted as the number of handover failures divided by the handover triggered numbers.

Although PPP is tractable in modeling the deployment of BSs, it is not suitable enough to capture the coupling between UEs and BSs in traffic hotspots, thus motivating the authors in \cite{Saha2018,Saha2019} to model the BSs and users in ultra-dense area using PCPs.
The performance of coverage probability for a typical device-to-device (D2D) receiver under two realistic scenarios was investigated in \cite{Afshang2016}. 
The analysis quantifies the best and worst performance of clustered D2D networks in terms of coverage probability and area spectral efficiency.
In order to capture the correlation between clustered BSs and users in high density area, the downlink coverage probability was discussed in cellular networks modeled by Thomas cluster process (TCP) and Mat\'{e}rn cluster process in \cite{Saha2017}.
In \cite{Afshang2018}, the performance of coverage probability and throughput was studied in heterogeneous networks (HetNets) under two association policies, the simulations demonstrated that with the increasing reusability of same resource block for small cells, the coverage probability decreases whereas throughput increases.

Base on the former research, it is well accepted that using random spatial point process to model the distribution of BSs is more realistic when comparing with the traditional hexagonal model \cite{ElSawy2017,Zhang2016} in ultra-dense networks.
In aforementioned papers, the distribution of SBSs is usually assumed to be same as the macro BSs (MBSs) to reduce the analytical and computational complexity.
But in the practical circumstances, especially in the urban area, the clustered SBSs or the users are usually non-homogeneously distributed.
For instance, in the city planning, several specific areas are chosen as the economic or educational centers. Those geographical centers usually have higher population density and communication demand compared with other areas \cite{YAydin2021}. 
And the movement trajectory of targets in those area tend to limit in a specific region. This phenomenon brings the edge effects into our study.
In that case, the practical location of SBSs among those centers are the superposition of non-uniform distributions.
In order to capture those features, we model the locations of SBSs clustered on the hotspots as PCP.
Moreover, in order to get the convincing handover analysis results, the selection of mobility model is quite important. 
It is essential to choose an appropriate mobility model to analyze how the movement of users affects the network performance. Normally the researchers select the random waypoint (RWP) model because of its tractability in modeling movement patterns of mobile nodes \cite{johnson1996dynamic}. 
The authors in \cite{Lin2013} provide an improved traceable mobility model which shows better performance in emerging cellular networks when compared with the classical RWP mobility model and the synthetic truncated Levy walk model. 
However, the aforementioned RWP model has some inherit disadvantages such as the nodes tend to concentrate on the center \cite{bettstetter2004stochastic} and the transition length scales with the size of the area.

In this paper, to overcome the difficulties that the traditional BSs deployment scheme can't apply in the hotspots areas, we develop a PCP based BSs deployment scheme to model the practical BSs distribution by considering the correlation between clustered BSs and users. 
Compared with \cite{Bao1,Bao2}, in our manuscript, we derive not only the handover rate, but also the handover failure rate and ping-pang rate to further investigate the handover performance in hotspots area. 
Moreover, in our PCP based model, the handover rate is also consisted of the probability that sojourn time for each user inside the ERB is larger than the threshold.
By considering the sojourn time of users inside the ERB circles, the prediction of handover rate is more accurate and reliable. 
In our revised manuscript, the location of the target BS is not restricted to any direction or any tier. 
The corresponding coordinate of target BS can be random point in the entire simulation region. 
Moreover, the average distance between different BSs under TCP is also presented, which is firstly proposed recently, to solve the handover problems in hotspots. 
Note that the density wave phenomenon (DWP) indicates that the nodes distributed in a finite area tend to concentrate on the center under RWP model \cite{Kerdsri2017,Xu2017}, which will bring the edge effects into our analysis. 
In this case, users have less probability to interact with those BSs near the edge areas. 
Therefore, we propose a modified RWP (MRWP) model to overcome the DWP and to increase the accuracy of handover decision in boundary region. 
Compared with the small cell coverage model in \cite{Xu2017}, where the target SBS is assumed in the horizontal direction of MBS, we adjust the coordinates of SBSs to ensure that the analysis of small cell coverage model is suitable for any SBS in the coverage area. In that case, the analysis of small cell coverage model is suitable for any SBS in the coverage area.

Incorporating the spatial coupling and the MRWP model, we first derive the closed-form expressions of handover rate, handover failure rate, and ping-pong rate. Then extensive simulations are conducted to validate the theoretical results and to analyze how the user mobility affects the handover performance in hotspot areas.

The main contributions of this paper are as follows:
\begin{itemize}
    \item Novel BSs deployment scheme and MRWP model are proposed. We develop a practical HetNets model for traffic hotspots area, where the locations of users and clustered SBSs are coupled in hotspots to capture the correlation. We also propose a MRWP model to eliminate the density wave
    phenomenon in hotspots area and to increase the accuracy of handover decision.
    \item New distances distributions and average distances are derived. Considering the method proposed in \cite{salahat2013simple} using a sum of finite exponential series to obtain the approximate expression of the integral of the first kind modified Bessel function, we derive the closed-form expressions of average distances from one SBS to other BSs, handover rate, handover failure rate, and ping-pong rate under TCP scenario.
    \item The accuracy of the theoretical results are verified through simulations. The effects and the tradeoff of system parameters, such as BS density, the scattering variance of TCP, thresholds of triggered time, and user mobility on the handover performances are studied to provide guidance for future network planning in ultra-dense HetNets.
\end{itemize}

\begin{table}[!t]
    \renewcommand{\arraystretch}{1.3}
    \setlength\tabcolsep{3pt}
    \centering\caption{Notations and Descriptions of Parameters}\label{table:parameter}
    \begin{tabular}{|p{6em}|p{23em}|}
        \hline
        \multicolumn{1}{|c|}{ \bfseries Parameters} & \multicolumn{1}{|c|}{ \bfseries Descriptions}\\
        \hline
        $\lambda_{\mathrm{M}}, \lambda_{\mathrm{S}}, \lambda_{\mathrm{S^\prime}}$ & Densities of MBSs, PPP-SBSs, and PCP-SBSs.
        \\
        \hline
        $\Phi_{\mathrm{M}}, \Phi_{\mathrm{S}}, \Phi_{\mathrm{S^\prime}}$ & Locations of MBSs, PPP-SBSs, and PCP-SBSs.
        \\
        \hline
        $\lambda_p$, $\sigma$ & Density of parent point and scattering standard variance of offspring point for $\Phi_{\mathrm{S^\prime}}$.
        \\
        \hline
        $P_i$, $G_i$ & Transmit power and antenna gain of BSs in the $i$-th tier.
        \\
        \hline
        $C_i$, $\alpha_i$ & Path-loss intercept and exponent in the $i$-th tier.
        \\
        \hline
        $B_i$ & Association bias of the $i$-th tier.
        \\
        \hline
        $L_k$ & Rayleigh distributed initial transition length in the $k$-th movement with parameter $\sigma_{\mathrm{RWP}}$.
        \\
        \hline
        $\mu_k$ & Bernoulli distributed indicator of transition length extension in the $k$-th movement with parameter $p_Z$.
        \\
        \hline
        $Z_k$ & Rayleigh distributed extended transition length in the $k$-th movement with parameter $\sigma_Z$.
        \\
        \hline
        $V_k$ & User velocity during the $k$-th movement.
        \\
        \hline
        $S_k$ & User pause time after the $k$-th movement.
        \\
        \hline
        $R_{ij}$ & Distance from a random point inside $\Phi_i$ to the nearest point inside $\Phi_j$, $i,j\in\{\mathrm{M},\mathrm{S},\mathrm{S^\prime}\}$.
        \\
        \hline
        $\bar{R}_{ij}$ & Average distance from a random point inside $\Phi_i$ to the nearest point inside $\Phi_j$, $i,j\in\{\mathrm{M},\mathrm{S},\mathrm{S^\prime}\}$.
        \\
        \hline
        $H_{\mathrm{t},ij}$ & Handover triggered rate from tier $i$ to tier $j$.
        \\
        \hline
        $H_{ij}$ & Handover rate from tier $i$ to tier $j$.
        \\
        \hline
        $H_{\mathrm{f},ij}$ & Handover failure rate from tier $i$ to tier $j$.
        \\
        \hline
        $H_{\mathrm{p},ij}$ & Ping-pong rate from tier $i$ to tier $j$.
        \\
        \hline
        $T$ & Time threshold of handover failure event.
        \\
        \hline
        $T_p$ & Time threshold of Ping-pong event.
        \\
        \hline
    \end{tabular}
\end{table}

\section{System Model}\label{sec:model}

\subsection{BS Deployment Model}\label{subsec:base station distribution}
\begin{figure}[!t]
    \centering
    \includegraphics[width=0.30\textwidth]{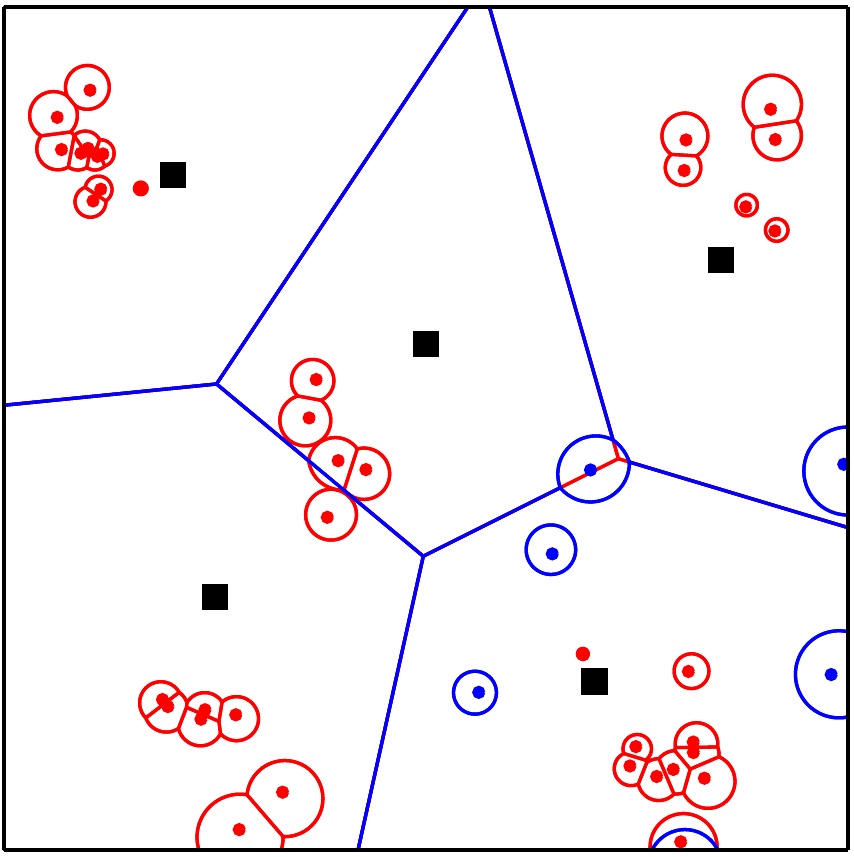}
    \caption{Illustrations of the BSs deployment model. The black square represents the location of MBS. The blue dot and the blue circle around it represent the location of SBS following PPP and its ERB, respectively. The red dot and the red circle around it represent the location of SBS following TCP and its ERB, respectively.
    }
    \label{fig1}
\end{figure}

We consider a three-tier HetNet consisting of uniformly deployed MBSs and SBSs in general area and clustered SBSs in hotspot area, as shown in Fig.~\ref{fig1}. 
The macro cells mainly provide global coverage and support the necessary signaling transmission \cite{Xu2017}, while the small cells are allocated with specific frequency to provide local area coverage and basic communication services.
Specifically, the locations of uniform MBSs and SBS are assumed to follow homogeneous PPPs $\Phi_{\mathrm{M}}$ and $\Phi_{\mathrm{S}}$ in the plane with density $\lambda_{\mathrm{M}}$ and $\lambda_{\mathrm{S}}$, respectively. 
Regarding the clustered SBSs, inspired by the fact that SBSs in the hotspot areas are usually geographically clustered, to capture the spatial correlation, we model these SBSs as a TCP $\Phi_{\mathrm{S}^\prime}$ generated from a parent point process $\Phi_p$. Here $\Phi_p$ is a homogeneous PPP with density $\lambda_p$, and the SBSs are independently and identically scattered around the cluster centers of $\Phi_p$ following a symmetric normal distribution with variance $\sigma^2$. 
Therefore, the probability density function (PDF) of distance $R$ from the offspring point to the cluster center can be expressed as
\begin{align}
    f_R\left( r \right) =\frac{1}{2\pi \sigma ^2}\exp \left( -\frac{r^2}{2\sigma ^2} \right) ,\quad r\ge 0.\label{eq:TCP_def}
\end{align}
Thus $\Phi_{\mathrm{S}^\prime}$ can be equivalently expressed as $\Phi _{\mathrm{S}^\prime}=\bigcup_{z\in \Phi _p}{B_z}$, where $B_z$ is the subset of $\Phi_{\mathrm{S}^\prime}$ generated from the parent point $z\in\Phi_p$.

\subsection{User Mobility Model}\label{subsec:mobility}
Based on the improved RWP model in \cite{Xu2017}, the user's $k$th movement can be represented by $(\bm{X}_{k-1}, \bm{X}_k,V_k,S_k)$, where $\bm{X}_{k-1}$ is the starting waypoint, $\bm{X}_{k}$ is the target waypoint, $V_{k}$ is the velocity, and $S_{k}$ is the pause time of user staying at $\bm{X}_{k}$.
Given the starting waypoint $\bm{X}_{k-1}$, the transition length $L_k=\left\| \bm{X}_k-\bm{X}_{k-1} \right\|$ follows a Rayleigh distribution with parameter $\sigma_{\mathrm{RWP}}$ and the direction is uniformly distributed in $[0,2\pi]$. 
By adjusting $\sigma_{\mathrm{RWP}}$, the average transition can be changed to model the movement patterns of users in infinite region. 
According to \cite{bettstetter2004stochastic}, the DWP will affect the accuracy of handover probability calculation.
It can be ignored when the simulation area is sufficient large, in which the number of points near the edge are small. 
However, as we described in Section~\ref{sec:introduction}, the trajectories of users in hotspots are usually clustered in finite regions. Thus the improved RWP model proposed in \cite{Xu2017}, which is designed for infinite planes, will not suitable for our limited finite regions.
In our manuscript, the handover analysis is conducted on a hotspot area, and the edge effect will thus bring difficulties to our handover performance analysis.
To minimize the impact of DWP, we propose the MRWP model by inducing a Rayleigh distributed extra distance $Z$ to extend the trajectory of each movement with a certain chance while keeping the direction unchanged.
Specifically, the transition length after extension in MRWP is formulated by 
\begin{align}
    L_{k}^{\prime}=L_k+\mu_k Z_k,\label{eq:modified User mobility}
\end{align}
where $\mu_k$ is a Bernoulli random variable with parameter $p_Z$ and the $Z_k$ is a Rayleigh random variable with PDF given by
\begin{align}
    f_{Z_k}\left( z \right) =\frac{z}{\sigma _{Z}^{2}}\exp \left( -\frac{z^2}{2\sigma _{Z}^{2}} \right),\quad z\ge 0.\label{eq:the pdf of random variable z}
\end{align}
Besides, we assume that the velocity $V_{k}$ and the pause time $S_{k}$ are independently and identically distributed with PDF $f_V\left( \cdot \right)$ and $f_S\left( \cdot \right)$, respectively.

\begin{figure}[!t]
    \centering
    \subfloat[Illustration of separated simulation areas]{\includegraphics[width=0.35\textwidth]{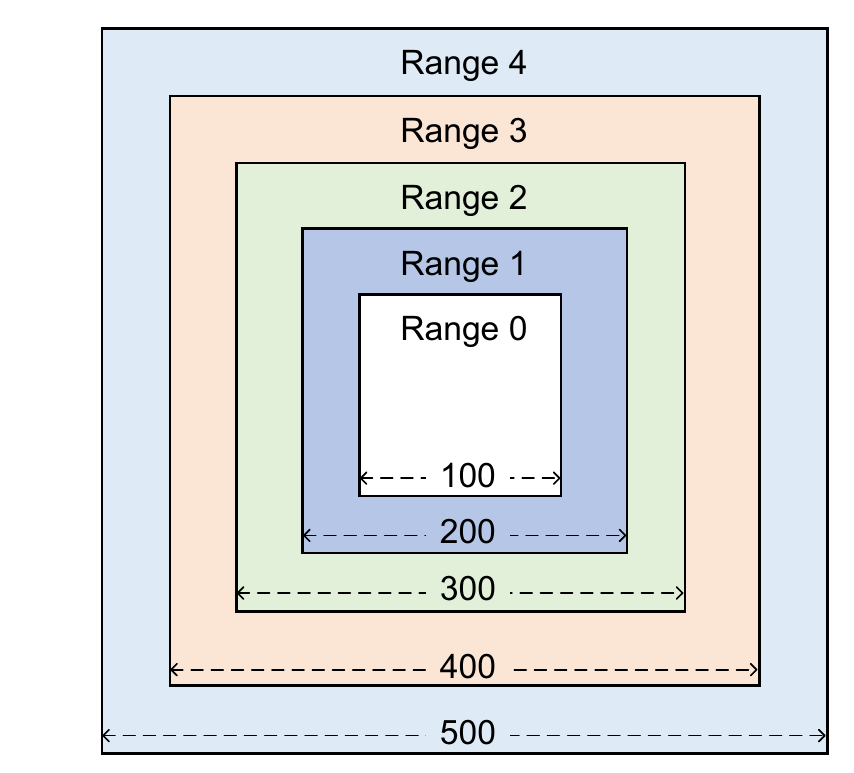}\label{fig2a}}
    \hfil
    \subfloat[The occurrence probability]{\includegraphics[width=0.38\textwidth]{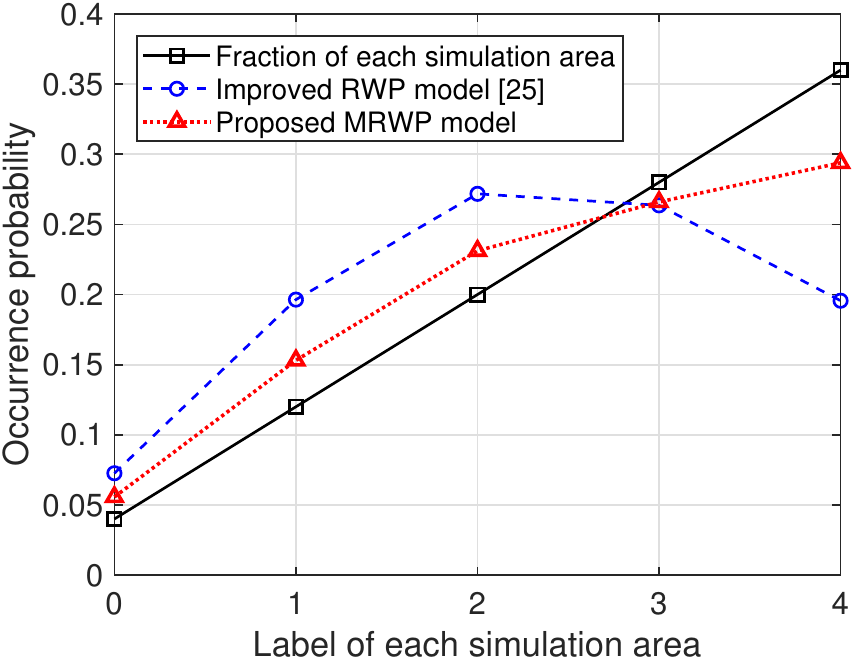}\label{fig2b}}
    \caption{The partition of simulation area and the occurrence probability for MRWP model and improved RWP model.}
    \label{fig2}
\end{figure}

To validate the effectiveness of MRWP model, we separate the simulation region into five sub-regions in Fig.~\ref{fig2a}. 
The occurrence probabilities of users in these regions under RWP model and MRWP model are presented in Fig.~\ref{fig2b}. 
Compared with RWP model, MRWP model results in higher occurrence probability in boundary region and more uniform waypoints in whole region, which shows that the handover analysis under MRWP model is more reliable and convincing.

\subsection{Channel Model and Association Strategy}\label{subsec:channel}
According to 3GPP standards, a handover procedure will be processed in 4 layers.
We consider the method in \cite{Xu2017} where prefect filtering are adopted in Layer-1 and Layer-4. 
Since the prefect filtering is deployed and dynamic resource allocation is assumed to average the fading and noise. Several related works also take similar assumptions in their analyses \cite{Xu2017,Saha2018}. The link experiences power-law path-loss.
The handover procedure usually takes the instantaneous reference signal received power (RSRP) as the indicator to initiate or terminate the process.
We assume that the BSs in $i$th tier have identical transmit power $P_{i}$ and antenna gain $G_{i}$, $i\in\{\mathrm{M}, \mathrm{S}, \mathrm{S}^\prime\}$.
The RSRP can be simplified to the downlink received signal strength (DL-RSS) as
\begin{align}
    \mathrm{RSS}_i\left( r \right) =B_iP_iG_iC_ir^{-\alpha _i},\label{eq:RSSI}
\end{align}
where $r$ is the distance from the typical terminal to the BS in tier $i$, $B_{i}$ represents association bias which aims at adjusting the coverage range of BS to achieve load balance, $C_{i}$ is the path-loss at reference distance, and $\alpha_{i}$ is the path-loss exponent. 
Along the mobility trajectory, the user will choose the BS which provides the strongest DL-RSS as the next serving BS. 
The handover boundary for each BS is then determined by the equal long-term biased DL-RSS boundary (ERB) \cite{LopezPerez2012a}.

\section{Handover Boundary and Distance distribution Analysis}\label{sec:boundary_distance}

In this section, we first determine the approximate expressions of handover boundary and handover failure boundary, and then derive the distribution of the distance between the BSs that the user is served by before and after handover. These metrics will be used to characterize the handover performance in traffic hotspot area in Section~\ref{sec:handover_analysis}.

\subsection{Handover Boundary Analysis}\label{subsec:boundary_analysis}
Consider a typical BS $X_i\in \Phi _i$ at origin and a target BS $X_j\in \Phi _j$, the ERB $\mathscr{B}_{X_i X_j}$ is comprised of the points at which the DL-RSSs received from $X_i$ and $X_j$ are identical, i.e.,
\begin{align}
    \mathscr{B} _{X_iX_j}=\left\{ X\in \mathbb{R} ^2 \mid \mathrm{RSS}_i\left( d_i \right) =\mathrm{RSS}_j\left( d_j \right) \right\},\label{eq:bound_circ_1}
\end{align}
where $d_i=\left\| X-X_i \right\|$ and $d_j=\left\| X-X_j \right\|$.
Recalling \eqref{eq:RSSI} and the Cartesian coordinates $X_i=\left( 0,0 \right)$, $X_j=\left( x_j,y_j \right)$, $X=\left( x,y \right)$, the condition in \eqref{eq:bound_circ_1} can be rewritten as
\begin{align}
    \xi_{ij}\left( x^2+y^2 \right) ^{\alpha_{ij}}-\left[ \left( x-x_j \right) ^2+\left( y-y_j \right) ^2 \right] =0,\label{eq:bound_circ_3}
\end{align}
where $\xi_{ij}=\left( B_jP_jG_jC_j/\left( B_iP_iG_iC_i \right) \right) ^{2/\alpha _j}$ and $\alpha_{ij}={\alpha _i}/{\alpha _j}$.
According to the 3GPP simulation methodology \cite{3gpp2013study}, path-loss exponents for MBS and SBS in the outdoor scenario are quite similar, i.e., $\alpha_{ij}\approx 1$. 
For analytical tractability, we approximate $f\left( x,y \right) =\left( x^2+y^2 \right) ^{\alpha_{ij}}$ with $\tilde{f}\left( x,y \right) =\lambda \left( x^2+y^2 \right)$ by minimizing $E_{\lambda}\left( x,y \right)$, where
\begin{align}
    E_{\lambda}\left( x,y \right) &=\left| f\left( x,y \right) -\hat{f}\left( x,y \right) \right|\notag
    \\
    &=\left| \left( x^2+y^2 \right) ^{\alpha_{ij}}-\lambda \left( x^2+y^2 \right) \right|.\label{eq:bound_circ_4}
\end{align}
Converting coordinates from Cartesian to polar, we can rewrite $E_{\lambda}\left( x,y \right)$ to
\begin{align}
    E_{\lambda}\left( r \right) =\left| r^{2\alpha_{ij}}-\lambda r^2 \right|,\label{eq:bound_circ_5}
\end{align}
where $r=\sqrt{x^2+y^2}$.
The optimal $\lambda$ can then be expressed as
\begin{align}
    \lambda^*=\underset{\lambda}{\mathrm{arg}\min}\int_0^{\sqrt{x_j^{2}+y_j^{2}}}{E_{\lambda}\left( r \right) \mathrm{d}r}.\label{eq:bound_circ_6}
\end{align}
The exact expression of $\lambda^*$ is given in the following lemma.

\begin{lemma}\label{lemma:lambda_optimal}
$\lambda^*$ is determined by $\lambda^*=\left( x_j^{2}+y_j^{2} \right) ^{\alpha_{ij}-1}$.
\end{lemma}
\begin{IEEEproof}
See Appendix~\ref{appendix:lambda_optimal}.
\end{IEEEproof}

Then, leveraging Lemma~\ref{lemma:lambda_optimal} and substituting $f\left( x,y \right)$ with $\tilde{f}\left( x,y|\lambda^* \right)$, the ERB in \eqref{eq:bound_circ_1} can be approximated by a circle $\mathcal{B} ( \tilde{X},\tilde{R} )$, where the center $\tilde{X}$ and radius $\tilde{R}$ are respectively given by
\begin{align}
    \tilde{X} &=\left( \frac{x_j}{1-\lambda ^*\xi _{ij}},\frac{y_j}{1-\lambda ^*\xi _{ij}} \right), \tilde{R} =\frac{\sqrt{\lambda ^*\xi _{ij}\left( x_{j}^{2}+y_{j}^{2} \right)}}{1-\lambda ^*\xi _{ij}}.\label{eq:handover_boundary}
\end{align}

\begin{figure}[!t]
    \centering
    \includegraphics[width=0.38\textwidth]{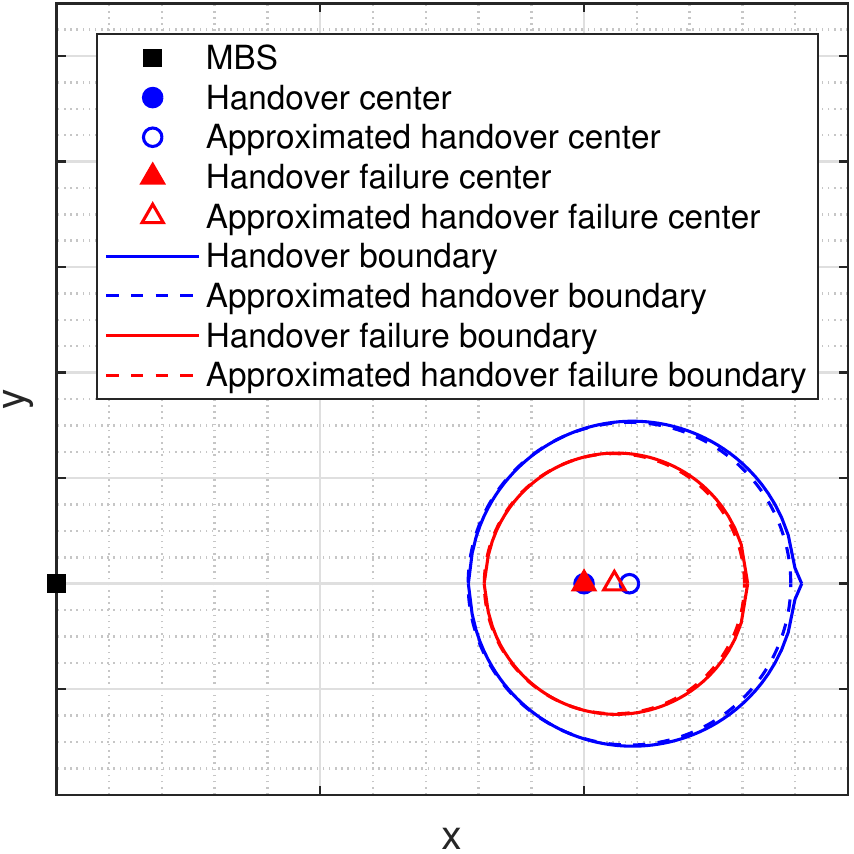}
    \caption{The approximate solution of handover circle boundary and handover failure circle boundary.}
    \label{fig3}
\end{figure}

The approximated closed-form circular equation of handover initial process is presented in \eqref{eq:handover_boundary}. 
Similarly, the handover failure boundary can be approximated by $\mathcal{B} ( \tilde{X}_{\mathrm{f}},\tilde{R}_{\mathrm{f}} )$ with 
\begin{align}
    \tilde{X}_{\mathrm{f}}=\left( \frac{x_j}{1-\lambda ^*\xi _{\mathrm{f},ij}},\frac{y_j}{1-\lambda ^*\xi _{\mathrm{f},ij}} \right),\tilde{R}_{\mathrm{f}}=\frac{\sqrt{\lambda ^*\xi _{\mathrm{f},ij}\left( x_{j}^{2}+y_{j}^{2} \right)}}{1-\lambda ^*\xi _{\mathrm{f},ij}},
\end{align}
where $\xi _{\mathrm{f},ij}=\xi _{ij}Q_{\mathrm{out}}^{2/\alpha _j}$ and $Q_{\mathrm{out}}<1$.
Once the UE moves into the handover triggered circle, the threshold time $T$ starts. If the UE steps into the handover failure circle within $T$, the handover process failed.
Based on the aforementioned analysis, the ERB functions of handover and handover failure are approximated by the circular equations as shown in Fig.~\ref{fig3}. Combining the MRWP model and handover boundary analysis, the handover rate and handover failure rate can be determined by the number of UEs' trajectory interacting with $\mathcal{B}( \tilde{X},\tilde{R} )$ and $\mathcal{B}( \tilde{X}_{\mathrm{f}},\tilde{R}_{\mathrm{f}} )$ in the unit time, respectively.

\subsection{Handover Strategy}\label{subsec:handover_strategy}
In this paper, by combing the MRWP model and BSs deployment under prefect filtering and dynamic resource allocation, the users can switch between different BSs according to DL-RSS.
The handover process is divided into two sub-processes, termed as handover triggered process and the sojourn time process. 
The handover event, handover failure event, and the ping-pang event are defined as follows:
\begin{itemize}
    \item Only if the user crosses the boundary of handover circle and stays inside handover ERB longer than $T$, a handover procedure is determined as successful.
    \item As for the handover failure process, when the user crosses the boundary of handover failure circle, in the meantime, the sojourn time between handover ERB and handover failure ERB is smaller than $T$, The handover procedure is determined as unsuccessful.
    \item For the ping-pong process, after the handover procedure is initiated, the typical user returns to the original BS service area before the sojourn time is greater than $T_p$ inside the target cell. In that case, a ping-pong procedure is determined as successful.
\end{itemize}
To be noted that each of the above handover events, including the handover event, handover failure
event, and the ping-pang event occurs on the premise of handover triggered event occurring. 

\subsection{Distance Distribution}\label{subsec:distance_pdf}
For the analysis of the average distances and handover performance in the proposed model, we need to get the distribution of those relevant distances from the two kind of SBSs to the MBS.
Let $R_{ij}$ denote the distances from $x_i\in\Phi_i$ to the nearest BS in $\Phi_j$. 
Since the coverage area is a circle according to Section~\ref{subsec:boundary_analysis}, we approximate $R_{ij}$ by its average value.
$R_{\mathrm{S^\prime M}}$ represents the distance from a typical SBS in $\Phi_{\mathrm{S}^\prime}$ to the nearest MBS in $\Phi_{\mathrm{M}}$.
 $R_{\mathrm{SM}}$ represents the distance from a typical SBS in $\Phi_{\mathrm{S}}$ to the nearest MBS in $\Phi_{\mathrm{M}}$. As for $R_{\mathrm{S^\prime S}}$, which represents the distance from a typical SBS in $\Phi_{\mathrm{S}^\prime}$ to the nearest SBS in $\Phi_{\mathrm{S}}$.

\subsubsection{$R_{\mathrm{SM}}$}
The PDF and CDF of $R_{\mathrm{SM}}$ are given by
\begin{align}
    &f_{R_{\mathrm{SM}}}\left(r\right) =2 \pi \lambda_{m} r \exp \left(-\pi \lambda_{m} r^{2}\right),
    \\
    &F_{R_{\mathrm{SM}}}\left(r\right) =1-\exp \left(-\pi \lambda_{m} r^{2}\right),
\end{align}
respectively.

\subsubsection{$R_{\mathrm{S^\prime S}}$}
Assuming the typical BS located at origin, the distribution of $R_{\mathrm{S^\prime S}}$ conditioned on the hotspot center location is given in the following theorem.
\begin{theorem}\label{theorem:minDistancePDF}
Conditioned on the distance from the hotspot center to the origin being $W_{\mathrm{S}}$, the conditional PDF and CDF of $R_{\mathrm{S^\prime S}}$ are
\begin{align}
    &f_{R_{\mathrm{S^\prime S}}|W_{\mathrm{S}}}\left( r|w \right) =\frac{r}{\sigma ^2}\exp \left( -\frac{r^2+w^2}{2\sigma ^2} \right) I_0\left( \frac{wr}{\sigma ^2} \right),\label{eq:R_SS_PDF}
    \\
    &F_{R_{\mathrm{S^\prime S}}|W_{\mathrm{S}}}\left( r|w \right) =1-Q_1\left( \frac{w}{\sigma ^2},\frac{r}{\sigma ^2} \right),\label{eq:R_SS_CDF}
\end{align}
respectively, where $f_{W_{\mathrm{S}}}\left( w \right) =2\pi \lambda _{\mathrm{S}}w\exp \left( -\pi \lambda _{\mathrm{S}}w^2 \right)$, $I_0$ is the modified Bessel function of the first kind, and $Q_{1}\left( \cdot \right)$ is the Marcum Q-function.
\end{theorem}
\begin{IEEEproof}
See Appendix~\ref{appendix:minDistancePDF}.
\end{IEEEproof}

\subsubsection{$R_{\mathrm{S^\prime M}}$}
As for $R_{\mathrm{S^\prime M}}$, which represents the distance from a typical SBS in $\Phi_{\mathrm{S}^\prime}$ to the nearest MBS in $\Phi_{\mathrm{M}}$.
Similar to $R_{\mathrm{S^\prime S}}$, the conditional PDF and CDF of $R_{\mathrm{S^\prime M}}$ can be expressed as
\begin{align}
    &f_{R_{\mathrm{S^\prime M}}|W_{\mathrm{M}}}\left(r|w \right)=\frac{r}{\sigma ^2}\exp \left( -\frac{r^2+w^2}{2\sigma ^2} \right) I_0\left( \frac{wr}{\sigma ^2}\right),
    \\
    &F_{R_{\mathrm{S^\prime M}}|W_{\mathrm{M}}}\left(r|w \right)=1-Q_1\left( \frac{w}{\sigma ^2},\frac{r}{\sigma ^2} \right),
\end{align}
where $f_{W_{\mathrm{M}}}\left( w \right) =2\pi \lambda _{\mathrm{M}}w\exp \left( -\pi \lambda _{\mathrm{M}}w^2 \right)$.

\subsection{Mean Distance}
The handover rate can be separated into handover triggered rate and the probability that the mobile terminal's sojourn time inside the target cell is larger than the threshold.
To obtain the handover triggered rate under different distances distributions, it is essential to calculate the average values of $R_{\mathrm{SM}}$, $R_{\mathrm{S^\prime S}}$, and $R_{\mathrm{S^\prime M}}$. 

\subsubsection{$\bar{R}_{\mathrm{SM}}$}
The average value of $R_{\mathrm{SM}}$ can be obtained by
\begin{align}
    \bar{R}_{\mathrm{SM}}=\mathbb{E} \left[ R_{\mathrm{SM}} \right] =\int_0^{\infty}{rf_{R_{\mathrm{SM}}}\left( r \right) \,\mathrm{d}r}=\frac{1}{2\sqrt{\lambda _{\mathrm{M}}}}.\label{eq:F_rw0_10}
\end{align}

\subsubsection{$\bar{R}_{\mathrm{S^\prime S}}$}

\begin{table}[!t]\scriptsize
    \renewcommand{\arraystretch}{1.3}
    \setlength\tabcolsep{2pt}
    \centering\caption{The values of $a_k$ and $b_k$ in \eqref{eq:F_rw0_11} for different intervals of $z$}\label{table:bessel_approx}
    \begin{tabular}{|*{8}{c|}}
        \hline
        \multicolumn{2}{|c|}{$0\leq z< 11.5$} & \multicolumn{2}{|c|}{$11.5\leq z< 20$} & \multicolumn{2}{|c|}{$20\leq z< 37.25$} & \multicolumn{2}{|c|}{$z\geq 37.25$} \\
        \hline
        $a_k$ & $b_k$ & $a_k$ & $b_k$ & $a_k$ & $b_k$ & $a_k$ & $b_k$\\
        \hline
        $0.1682$ & $0.7536$ & $0.2667$ & $0.4710$ & $0.1121$ & $0.9807$ & $2.4e^{-9}$ & $1.144$\\
        \hline
        $0.1472$ & $0.9736$ & $0.4916$ & $-163.4$ & $0.1055$ & $0.8672$ & $0.0675$ & $0.995$\\
        \hline
        $0.4450$ & $-0.715$ & $0.1110$ & $0.9852$ & $-1.8e^{-4}$ & $1.0795$ & $0.0547$ & $0.567$\\
        \hline
        $0.2382$ & $0.2343$ & $0.1304$ & $0.8554$ & $0.0033$ & $1.0385$ & $0.0787$ & $0.946$\\
        \hline
    \end{tabular}
\end{table}

The average value of $R_{\mathrm{S^\prime S}}$ can be expressed as
\begin{align}
    \bar{R}_{\mathrm{S^\prime S}} &=\mathbb{E} \left[ R_{\mathrm{S^\prime S}} \right] =\mathbb{E} _{W_\mathrm{S}}\left[ \int_0^{\infty}{rf_{R_{\mathrm{S^\prime S}}|W_{\mathrm{S}}}\left( r\right) \mathrm{d}r} \right].\label{eq:F_rw0_001}
\end{align}
Substituting \eqref{eq:R_SS_PDF} into \eqref{eq:F_rw0_001} yields
\begin{align}
    \bar{R}_{\mathrm{S^\prime S}}&=\int_0^{\infty}\int_0^{\infty}\frac{2\pi \lambda _\mathrm{S}wr^2}{\sigma ^2}\exp \left( -\frac{r^2+w^2}{2\sigma ^2}-\pi \lambda _\mathrm{S}w^2 \right)\notag
    \\
    &\qquad\qquad\qquad\qquad\times I_0\left( \frac{wr}{\sigma ^2} \right) \mathrm{d}r\mathrm{d}w.\label{eq:F_rw0_11}
\end{align}
To obtain the closed-form expression of $\bar{R}_{\mathrm{S^\prime S}}$, we approximate $I_0\left( z \right)$ by the sum of a finite exponential series \cite{salahat2013simple}, i.e.,
\begin{align}
    I_0\left( z \right) \approx \sum_{k=1}^4{a_ke^{b_kz}},\label{eq:F_rw0_12}
\end{align}
where $a_k$ and $b_k$ are presented in Table~\ref{table:bessel_approx}. 
Therefore, substituting \eqref{eq:F_rw0_12} into \eqref{eq:F_rw0_11}, $\bar{R}_{\mathrm{S^\prime S}}$ can be simplified to
\begin{align}
    \bar{R}_{\mathrm{S^\prime S}}=\sum_{k=1}^4{\frac{2\pi \lambda _\mathrm{S} a_k}{\sigma ^2}\int_0^{\infty}{we^{-\left( \frac{1}{2\sigma ^2}+\pi \lambda _\mathrm{S} \right) w^2}F_k\left( w \right) \mathrm{d}w}},\label{eq:meanvalueRS22S}
\end{align}
where
\begin{align}
    F_k\left( w \right) =\int_0^{\infty}{r^2\exp \left( -\frac{r^2}{2\sigma ^2}+\frac{b_kwr}{\sigma ^2} \right) \mathrm{d}r}.\label{eq:F_W02}
\end{align}
By invoking \cite[3.462.7]{GradRyzh14BK}, $F_k\left( w \right)$ can be evaluated by
\begin{align}
    F_k\left( w \right) &=\sigma ^2 b_kw+\sqrt{\frac{\pi}{2}}\left( \sigma b_{k}^{2}w^2+\sigma ^3 \right) e^{\frac{b_{k}^{2}w^2}{2\sigma ^2}}\notag
    \\
    &\qquad\qquad\qquad\times\left[ 1+\mathrm{erf}\left( \frac{b_kw}{\sqrt{2}\sigma} \right) \right],\label{eq:proofF_w0}
\end{align}
where $\mathrm{erf}\left( x \right) =\frac{2}{\sqrt{\pi}}\int_0^x{\exp \left( -y^2 \right) \mathrm{d}y}$ is the error function.


Since $\mathrm{erf}\left( x \right)< 1$, $x\geq 0$, we obtain an upper bound of $F_k\left( w \right)$ as follows
\begin{align}
    F_{k}^{\left( \mathrm{ub} \right)}\left( w \right) =\sqrt{2\pi} \left( \sigma b_{k}^{2}w^2+\sigma ^3 \right) e^{\frac{b_{k}^{2}w^2}{2\sigma ^2}}+\sigma ^2 b_kw.
\end{align}
Substituting $F_{k}^{\left( \mathrm{ub} \right)}\left( w \right)$ into \eqref{eq:meanvalueRS22S}, we obtain an upper bound of $\bar{R}_{\mathrm{S^\prime S}}$ as follows
\begin{align}
    \bar{R}_{\mathrm{S^\prime S}}^{\left( \mathrm{ub} \right)} &=\sum_{k=1}^4 \frac{2\pi \lambda _\mathrm{S} a_k}{\sigma ^2}\int_0^{\infty} \sigma ^2 b_kw^2e^{-\left( \pi \lambda _\mathrm{S}+\frac{1}{2\sigma ^2} \right) w^2}\notag
    \\
    &\qquad +\sqrt{2\pi}w\left( \sigma b_{k}^{2}w^2+\sigma ^3 \right) e^{-\left( \pi \lambda _\mathrm{S}+\frac{1-b_{k}^{2}}{2\sigma ^2} \right) w^2} \mathrm{d}w\notag
    \\
    &=\sum_{k=1}^4 \sqrt{2\pi}q_{\mathrm{S}}\sigma a_k\left[ \frac{2}{\left( 2q_{\mathrm{S}}+1-b_{k}^{2} \right)} \right.\notag
    \\
    &\qquad\left.+\frac{b_k}{\left( 2q_{\mathrm{S}}+1 \right) ^{3/2}} +\frac{4b_{k}^{2}}{\left( 2q_{\mathrm{S}}+1-b_{k}^{2} \right) ^2} \right],\label{eq:up_boundary_of_RS22S}
\end{align}
where $q_{\mathrm{S}}=\pi \lambda_{\mathrm{S}}\sigma ^2$.

\subsubsection{$\bar{R}_{\mathrm{S^\prime M}}$}
Similarly, the average value of $\bar{R}_{\mathrm{S^\prime M}}$ is upper bounded by
\begin{align}
    \bar{R}_{\mathrm{S^\prime M}}^{\left( \mathrm{ub} \right)} &=\sqrt{2\pi}q_{\mathrm{M}}\sigma \sum_{k=1}^4 a_k\left[ \frac{2}{\left( 2q_{\mathrm{M}}+1-b_{k}^{2} \right)} \right.\notag
    \\
    &\qquad\quad\left.+\frac{b_k}{\left( 2q_{\mathrm{M}}+1 \right) ^{3/2}} +\frac{4b_{k}^{2}}{\left( 2q_{\mathrm{M}}+1-b_{k}^{2} \right) ^2} \right],\label{eq:up_boundary_of_RSppM_U}
\end{align}
where $q_{\mathrm{M}}=\pi \lambda_{\mathrm{M}}\sigma ^2$.

\section{Handover Performance Analysis}\label{sec:handover_analysis}

In this section, we give the definitions of handover rate, handover failure rate, and ping-pong rate. By formulating different handover procedures into several sub-processes, the theoretical expressions of these handover metrics under different BS densities and scattering variances of TCP are obtained.

\subsection{Handover Rate}\label{subsec:handover_rate}

\begin{figure}[!t]
    \centering
    \includegraphics[width=0.45\textwidth]{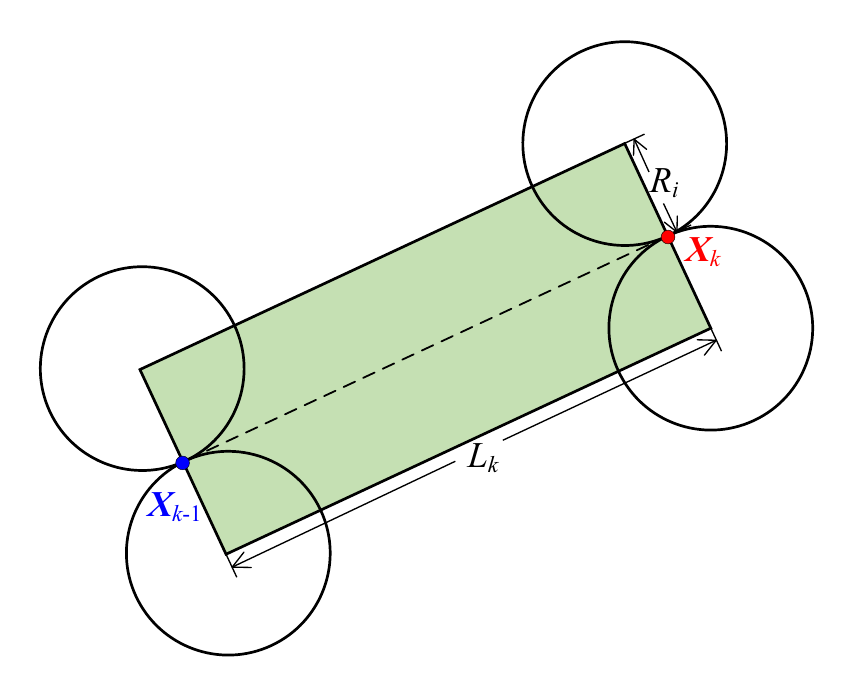}
    \caption{Illustration of the region in which the BS interacts with users' trajectory.}
    \label{fig4}
\end{figure}

According to the MRWP model, the trajectory of UE is divided into finite segments by setting the start waypoints and target waypoints.
The handover triggered rate is defined as the number of crosspoints of each UE's trajectory interacting with the ERB circle in the unit time.
According to Section~\ref{subsec:mobility}, the length of each movement is determined by $L_{k}^{\prime}=L_k+\mu_k Z_k$. During each movement, the handover event is triggered when the trajectory of the user interacts with the ERB circle, i.e., the distance from the nearest target BS to the trajectory of this movement is smaller than $\bar{R}_\ell$, $\ell\in \left\{ \mathrm{SM},\mathrm{S^\prime M},\mathrm{S^\prime S} \right\}$.
Since each trajectory and the radius of ERBs are finite, the BSs have to stay in the specific area to ensure that their ERBs can intersect with the user's trajectory.
We approximate this specific area as a rectangle with length $L_k$ and width $\bar{R}_\ell$.
As shown in Fig.~\ref{fig4}, the length is user's trajectory and the width is the distance from the center of the nearest ERB circle $\mathcal{B}( \tilde{X},\tilde{R})$ to the user's trajectory. 
Thus the specific area for each target BS can be expressed as $2L_k\bar{R}_\ell$. On the entire coverage, the probability that each target BS can be reached by user's trajectory is $\frac{2L_k\bar{R}_\ell}{\left| \mathcal{S} \right|}$.
Thus the \emph{handover triggered rate of one target BS} in the entire area can be expressed as
\begin{align}
    H_{\mathrm{t},\ell}^{\left( \mathrm{single} \right)} =\frac{1}{\left| \mathcal{S} \right|}\int_0^{\infty}{\int_0^{\infty}{2lrf_{L_k}\left( l \right) f_{R_\ell}\left( r \right) \mathrm{d}l}\mathrm{d}r},\label{eq:Handover_triggered_rate}
\end{align}
where $\mathcal{S}$ is the entire area and $\ell\in \left\{ \mathrm{SM},\mathrm{S^\prime M},\mathrm{S^\prime S} \right\}$.
Leveraging the independence between $L_k$ and $R_\ell$ and substituting \eqref{eq:handover_boundary} into \eqref{eq:Handover_triggered_rate} yield
\begin{align}
    H_{\mathrm{t},\ell}^{\left( \mathrm{single} \right)} &=\frac{2}{\left| \mathcal{S} \right|}\frac{\sqrt{\lambda ^*\xi}}{1-\lambda ^*\xi}\int_0^{\infty}{lf_{L_k}\left( l \right) \mathrm{d}l}\int_0^{\infty}{rf_{R_\ell}\left( r \right) \mathrm{d}r}\notag
    \\
    &=\frac{2}{\left| \mathcal{S} \right|}\frac{\sqrt{\lambda ^*\xi}}{1-\lambda ^*\xi}\mathbb{E} \left[ L_k \right] \mathbb{E} \left[ R_\ell \right],\label{eq:Handover_triggered_rate2}
\end{align}
which can be further simplified by using $\mathbb{E}\left[ L_k \right] =\frac{1}{2\sqrt{\sigma_{\mathrm{RWP}}}}+\sqrt{\frac{\pi}{2}}\sigma _Z$.
Considering there are $N_\mathrm{bs}$ target BSs in entire area and users move with constant velocity, we obtain the \emph{handover triggered rate} as follows
\begin{align}
    H_{\mathrm{t},\ell} &=\frac{\mathbb{E} \left[ N_{\mathrm{bs}} \right] H_{\mathrm{t},\ell}^{\left( \mathrm{single} \right)}}{\frac{\mathbb{E} \left[ L_k \right]}{\mathbb{E} \left[ V_k \right]}+\mathbb{E} \left[ S_k \right]}\notag
    \\
    &=\frac{2}{\left| \mathcal{S} \right|}\frac{\sqrt{\lambda ^*\xi}}{1-\lambda ^*\xi}\frac{\mathbb{E} \left[ N_{\mathrm{bs}} \right] \mathbb{E} \left[ R_\ell \right]}{\frac{1}{\mathbb{E} \left[ V_k \right]}+\frac{\mathbb{E} \left[ S_k \right]}{\mathbb{E} \left[ L_k \right]}}.\label{eq:7}
\end{align}

It is notable that the handover procedure may not be eventually executed when it is successfully triggered. 
From \eqref{eq:7} we can see that, if UEs do not pause in the entire movement, i.e., $\mathbb{E} \left[ S_k \right] =0$\cite{Xu2017,Lin2013}, then MRWP model has no impact on the handover triggered rate. 
Therefore, to exploit the impact of MRWP model on the handover performance, we assume a non-zero constant pause time in this paper. 

Once the UE moves into the ERB circle, the handover procedure is initiated.
According to \cite{Xu2017}, the \emph{average sojourn time} is formulated by
\begin{align}
    S_\ell=\frac{\pi R_\ell\sqrt{\lambda ^*\xi}}{2V\left( 1-\lambda ^*\xi \right)}.\label{eq:sojorun_time}
\end{align}
Then the probability that the UE stays in the ERB circle longer than the threshold $T$ is given by
\begin{align}
    &\mathbb{P} \left( S_{\ell}\ge T \right)\notag
    \\
    &=
    \begin{cases}
        \displaystyle \exp \left( -\frac{4\lambda _{\mathrm{M}}V^2T^2\left( 1-\lambda ^*\xi \right) ^2}{\pi \lambda ^*\xi} \right), & \ell =\mathrm{SM},\\
        \displaystyle Q_1\left( \frac{1}{2\sigma \sqrt{\lambda _{\mathrm{S}}}},\frac{2TV\left( 1-\lambda ^*\xi \right)}{\pi \sigma \sqrt{\lambda ^*\xi}} \right), & \ell =\mathrm{S^\prime S},\\
        \displaystyle Q_1\left( \frac{1}{2\sigma \sqrt{\lambda _{\mathrm{M}}}},\frac{2TV\left( 1-\lambda ^*\xi \right)}{\pi \sigma \sqrt{\lambda ^*\xi}} \right), & \ell =\mathrm{S^\prime M},\\
    \end{cases}\label{eq:sojorun_time_T}
\end{align}
respectively. 
Finally, as has been discussed in Section~\ref{subsec:handover_strategy}, the \emph{handover rate} is determined by
\begin{align}
    H_\ell=H_{\mathrm{t},\ell} \mathbb{P} \left( S_\ell\ge T \right).
\end{align}

\subsection{Handover Failure Rate}\label{subsec:handover_failure_rate}
The handover procedure is determined to be failed when the received SINR is smaller than the $Q_\mathrm{out}$ or the sojourn time is smaller than $T$.
The handover failure rate, which is identical to the handover failure numbers divided by the handover triggered numbers in a unit time, can be expressed as 
\begin{align}
    H_{\mathrm{f},\ell}=\frac{H_{\mathrm{f},\mathrm{t},\ell}\mathbb{P} \left( S_{\mathrm{f},\ell}\le T \right)}{H_{\mathrm{t},\ell}},
\end{align}
where $H_{\mathrm{f},\mathrm{t},\ell}$ is the handover failure triggered rate, and $S_{\mathrm{f},\ell}$ is the sojourn time for the typical UE between handover boundary and handover failure boundary.
Similar to \eqref{eq:7}, $H_{\mathrm{f},\mathrm{t},\ell}$ can be expressed as
\begin{align}
    H_{\mathrm{f},\mathrm{t},{\ell}} &=\frac{2}{\left| \mathcal{S} \right|}\frac{\sqrt{\lambda ^*\xi_{\mathrm{f}}}}{1-\lambda ^*\xi_{\mathrm{f}}} \frac{\mathbb{E} \left[ N_{\mathrm{bs}} \right] \mathbb{E} \left[ R_\ell \right]}{\frac{1}{\mathbb{E} \left[ V_k \right]}+\frac{\mathbb{E} \left[ S_k \right]}{\mathbb{E} \left[ L_k \right]}}.
\end{align}
And $\mathbb{P} \left( S_{\mathrm{f},\ell}\le T \right)$ can be expressed as
\begin{align}
    &\mathbb{P} \left( S_{\mathrm{f},\ell}\le T \right)\notag
    \\
    &=
    \begin{cases}
        \displaystyle 1-\exp \left( -\frac{4\lambda _{\mathrm{M}}V^2T^2\left( 1-\lambda ^*{\xi_{\mathrm{f}}} \right) ^2}{\pi \lambda ^*{\xi_{\mathrm{f}}}} \right), & \ell =\mathrm{SM},\\
        \displaystyle 1-Q_1\left( \frac{1}{2\sigma \sqrt{\lambda _{\mathrm{S}}}},\frac{2TV\left( 1-\lambda ^*{\xi_{\mathrm{f}}} \right)}{\pi \sigma \sqrt{\lambda ^*{\xi_{\mathrm{f}}}}} \right), & \ell =\mathrm{S^\prime S},\\
        \displaystyle 1-Q_1\left( \frac{1}{2\sigma \sqrt{\lambda _{\mathrm{M}}}},\frac{2TV\left( 1-\lambda ^*{\xi_{\mathrm{f}}} \right)}{\pi \sigma \sqrt{\lambda ^*{\xi_{\mathrm{f}}}}} \right), & \ell =\mathrm{S^\prime M}.\\
    \end{cases}\label{eq:S_f_ell}
\end{align}

\subsection{Ping-pong Rate}\label{subsec:pingpang_rate}
The ping-pong rate, which can be defined as the probability that after the handover procedure is initiated, the typical user returns back to the original BS's serving region in the time period $T_{\mathrm{p}}$. According to \cite{Xu2017}, the ping-pong rate $H_{\mathrm{p},\ell}$ is formulated by
\begin{align}
    H_{\mathrm{p},\ell}=H_{\mathrm{t},\ell}\left[ \mathbb{P} \left( S_\ell\le T_{\mathrm{p}} \right) -\mathbb{P} \left( S_\ell\le T \right) \right].
\end{align}
According to \eqref{eq:sojorun_time_T}, $H_{\mathrm{p},\ell}$ can be expressed as
\begin{align}
    H_{\mathrm{p},\ell}=\begin{cases}
        \displaystyle H_{t_{\mathrm{SM}}}\left[ \exp \left( -\frac{4\lambda _{\mathrm{M}}V^2T^2\left( 1-\lambda ^*\xi \right) ^2}{\pi \lambda ^*\xi} \right) \right.\\
        \displaystyle\quad \left. -\exp \left( -\frac{4\lambda _{\mathrm{M}}V^2T_{p}^{2}\left( 1-\lambda ^*\xi _{\mathrm{f}} \right) ^2}{\pi \lambda ^*\xi _{\mathrm{f}}} \right) \right], &\!\!\! \ell =\mathrm{SM},\\
        \displaystyle H_{t_{\mathrm{S^\prime S}}}\left[ Q_1\left( \frac{1}{2\sigma \sqrt{\lambda _{\mathrm{S}}}},\frac{2TV\left( 1-\lambda ^*\xi \right)}{\pi \sigma \sqrt{\lambda ^*\xi}} \right) \right.\\
        \displaystyle\quad \left. -Q_1\left( \frac{1}{2\sigma \sqrt{\lambda _{\mathrm{S}}}},\frac{2T_pV\left( 1-\lambda ^*\xi _{\mathrm{f}} \right)}{\pi \sigma \sqrt{\lambda ^*\xi _{\mathrm{f}}}} \right) \right], &\!\!\! \ell =\mathrm{S^\prime S},\\
        \displaystyle H_{t_{\mathrm{S^\prime M}}}\left[ Q_1\left( \frac{1}{2\sigma \sqrt{\lambda _{\mathrm{M}}}},\frac{2TV\left( 1-\lambda ^*\xi \right)}{\pi \sigma \sqrt{\lambda ^*\xi}} \right) \right.\\
        \displaystyle\quad \left. -Q_1\left( \frac{1}{2\sigma \sqrt{\lambda _{\mathrm{M}}}},\frac{2T_pV\left( 1-\lambda ^*\xi _{\mathrm{f}} \right)}{\pi \sigma \sqrt{\lambda ^*\xi _{\mathrm{f}}}} \right) \right], &\!\!\! \ell =\mathrm{S^\prime M}.\\
    \end{cases}
\end{align}

\section{Numerical Results and Discussions}\label{sec:simulations}

In this section, we verify the analytical results of the handover performance which have been derived in the above sections.
In this paper, we focus on the cross-tier handovers between the two tiers of SBSs. The cross-tier handovers between MBSs and SBSs following PPP distribution, and those handovers occur in the same MBSs and SBSs ties have been investigated in the previous papers \cite{Bi2019, Stamou2019,Bao1, Bao2, Zhang2019}. 
The cross-tier handovers between MBSs and SBSs following PCP distribution share the same analytical method in our manuscript. As for the same tier, those handovers occur in the MBSs and SBSs following PPP with different densities share the same analytical method with \cite{Xu2017}.
The simulation area is set as a $5\,\text{km}\times 5\,\text{km}$ square where three-tiers dense BSs are deployed.
According to \cite{access2010further}, the path-loss of macro cells and small cells at distance $r$ are $128.1 + 37.6 \log_{10}\left(r\right)$~dB and $140.7 + 36.7 \log_{10}\left(r\right)$~dB, respectively.
The parameters of different cells are set as $P_\mathrm{M}=46$~dBm, $P_\mathrm{S}=P_\mathrm{S^\prime}=30$~dBm, $G_\mathrm{M}=14$~dBi, $G_\mathrm{S}=G_\mathrm{S^\prime}=5$~dBi, $B_\mathrm{M}=0$~dB, and $B_\mathrm{S}=B_\mathrm{S^\prime}=4$~dB. And $Q_\mathrm{out}$ is set as $-3$~dB.
For dense deployments, we assume $\lambda _\mathrm{S}=10\lambda _\mathrm{M}=10\lambda _p$.

\begin{figure}[!t]
    \centering
    \includegraphics[width=0.42\textwidth]{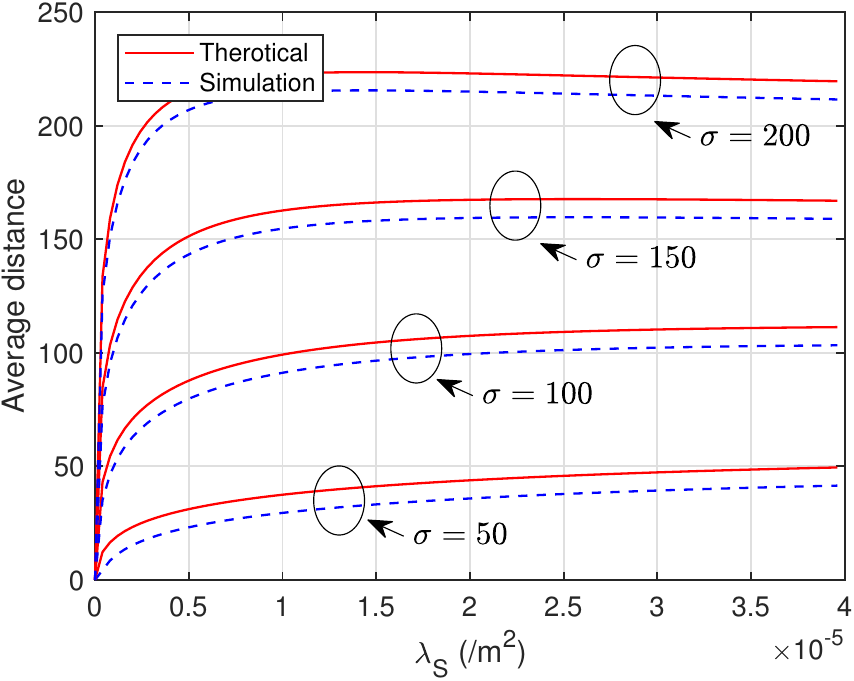}
    \caption{The average distance from the PCP distributed SBSs to the PPP distributed SBSs as a function of SBS density $\lambda_\mathrm{S}$.}
    \label{fig5}
\end{figure}

Fig.~\ref{fig5} shows the average distance $R_{\mathrm{S^\prime S}}$ as a function of $\lambda_\mathrm{S}$ under different $\sigma$.
As shown in Fig.~\ref{fig5}, the average value of $R_{\mathrm{S^\prime S}}$ increases with $\lambda _\mathrm{S}$ at first, and then tends to a stable value which is affected by $\sigma$.
It indicates that in the PCP HetNets, the average $R_{\mathrm{S^\prime S}}$ is mainly effected by $\sigma$ and $\lambda _\mathrm{S}$ has only limited influence when $\sigma$ is much larger than $\lambda _\mathrm{S}$.
This phenomenon is different from the results derived in PPP-based models.
According to Section~\ref{subsec:mobility}, the nodes in MRWP model tend to distribute in edge areas when the node density is small. Thus the average $R_{\mathrm{S^\prime S}}$ will increase with $\lambda_\mathrm{S}$ at the beginning.

\begin{figure}[!t]
    \centering
    \subfloat[The impact of PPP distributed SBS density $\lambda_\mathrm{S}$ when $\sigma=150$.]{\includegraphics[width=0.42\textwidth]{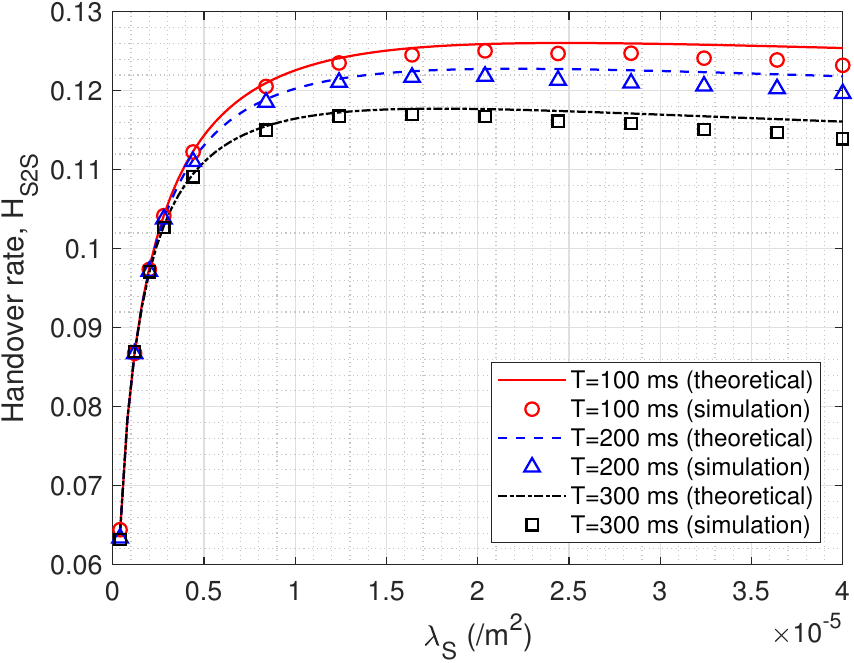}\label{fig6a}}
    \hfil
    \subfloat[The impact of PCP distributed SBS standard variance $\sigma$ when $\lambda_\mathrm{S}=2\times{10}^{-5}\text{/m}^{2}$.]{\includegraphics[width=0.42\textwidth]{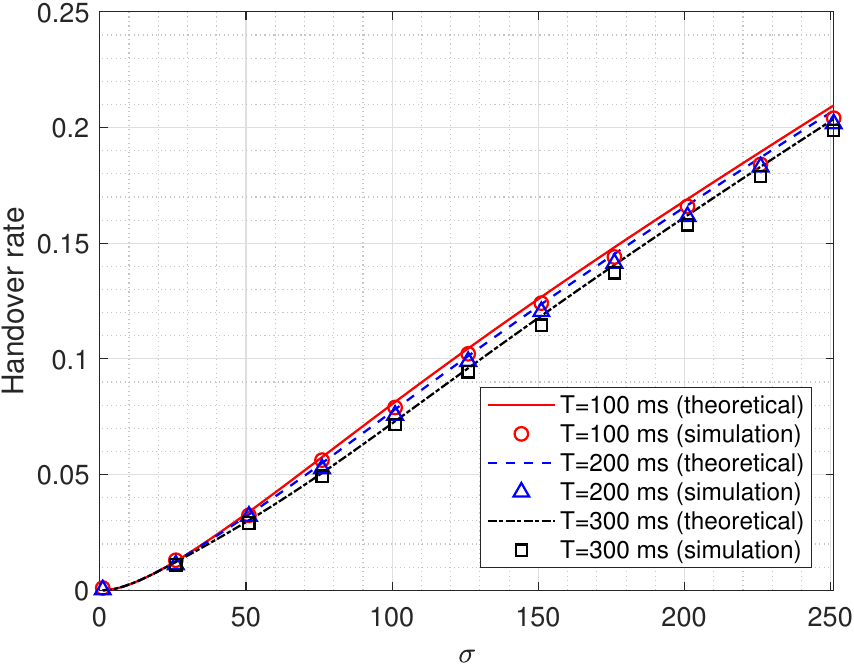}\label{fig6b}}
    \caption{The handover rate as a function of BS density and scattering standard variance at different sojourn time threholds $T$ when $P_\mathrm{S^\prime}=30$\,dBm and $V=60$\,km/h.}
    \label{fig6}
\end{figure}

Fig.~\ref{fig6} shows the handover rate as a function of $\lambda_\mathrm{S}$ and $\sigma$ under different thresholds $T$.
From Fig.~\ref{fig6a}, we can see that the handover rate follows the same trend with the average value of $R_{\mathrm{S^\prime S}}$.
The handover rate is consist of handover trigger rate and the probability that the handover UE's sojourn time inside the small cell is larger than the threshold $T$.
Because the handover triggered rate is approximately proportional to $R_{\mathrm{S^\prime S}}$ and $R_{\mathrm{S^\prime S}}$ is mainly affected by $\sigma$, the corresponding radius of each ERB will remain as a stable value with the increase of $\lambda _\mathrm{S}$ when $\sigma$ is a constant.
When the UE moves into the ERB circle, the average transmission time is more likely to be smaller than a larger $T$, which is obvious because the relative radius of the ERB is the function of $R_{\mathrm{S^\prime S}}$ and the average transmission time share the same trend with $R_{\mathrm{S^\prime S}}$.
The increasing threshold $T$ will lead to a lower handover rate as shown in Fig.~\ref{fig6a}.
In Fig.~\ref{fig6b}, the handover rate monotonically increases with $\sigma$.
According to \eqref{eq:handover_boundary} and the results in Fig.~\ref{fig5}, we conclude that the radius of ERB monotonically increases with average distance $R_{\mathrm{S^\prime S}}$.
Since the average distance $R_{\mathrm{S^\prime S}}$ is proportional to $\sigma$, the radius of ERB will also rise with the increasing $\sigma$.
In that case, the rising radius of ERB will result in higher coverage in each cluster.
Since the sojourn time for the user rises in a larger coverage, $\mathbb{P}(S_{\mathrm{S^\prime S}}\geq T)$ grows with the increase of $\sigma$, thus leading to a higher handover rate.

\begin{figure}[!t]
    \centering
    \subfloat[The impact of velocity $V$ when $P_\mathrm{S^\prime}=30$\,dBm.]{\includegraphics[width=0.42\textwidth]{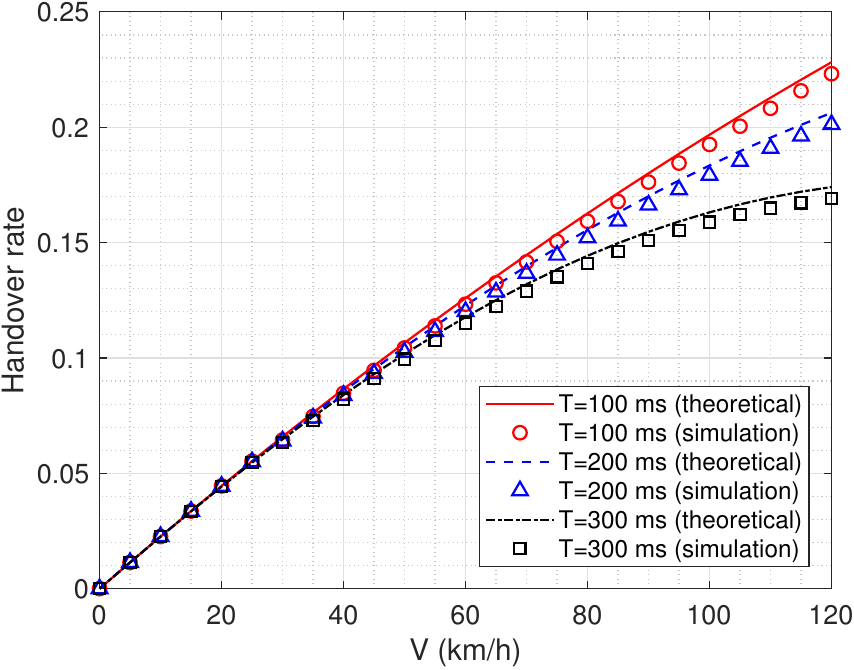}\label{fig7a}}
    \hfil
    \subfloat[The impact of transmit power $P_\mathrm{S^\prime}$ when $V=60$\,km/h.]{\includegraphics[width=0.42\textwidth]{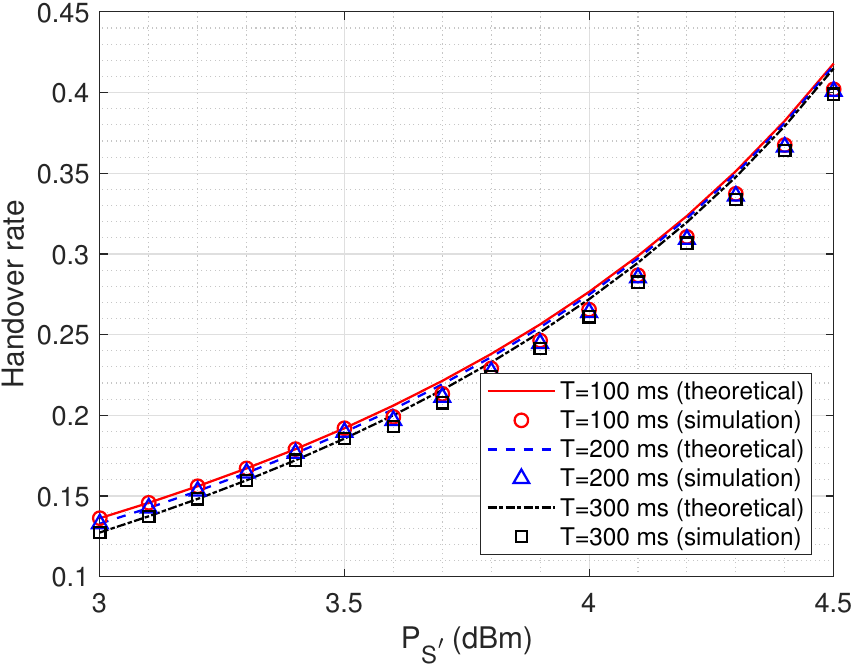}\label{fig7b}}
    \caption{The handover rate as a function of velocity $V$ and transmit power $P_\mathrm{S^\prime}$ at different sojourn time thresholds $T$ when $\lambda_\mathrm{S}=2\times{10}^{-5}\text{/m}^2$ and $\sigma=150$.}
    \label{fig7}
\end{figure}

The velocity of users will also influence the handover performance.
As shown in Fig.~\ref{fig7a}, the handover rate rises with the increasing $V$, which is obvious since the handover triggered rate is proportional to $V$ according to \eqref{eq:7}. When average distance $\bar{R}_\ell$ is unchanged, $\mathbb{P} \left( S_\ell\ge T \right)$ will decrease with the increasing $V$.
However, the trend of the decrease is nonlinear. As we can see in Fig.~\ref{fig7a}, with the increasing $V$, the handover rate shows a trend from rising to declining gradually, which can be explained by the fact the $\mathbb{P} \left( S_\ell\ge T \right)$ has reached the nonlinear part.
Fig.~\ref{fig7b} presents how the transmit power of typical BS influences the handover performance. 
When the transmit power $P_\ell$ of target BS is assigned with a stable value, the handover rate rises with the increasing transmit power $P_\mathrm{S^\prime}$ of typical BS. 
As shown in Section~\ref{subsec:boundary_analysis}, the radius of ERB circle is determined by the ratio of $P_\mathrm{S^\prime}$ and $P_\ell$. With the increasing $P_\mathrm{S^\prime}$, the radius of ERB circle will rise. From the simulations analysis, the higher $P_\mathrm{S^\prime}$ brings larger radius of ERB circle, which means the specific area in Section~\ref{subsec:handover_rate} gets larger. With unchanged simulation region, the ratio of specific area and the whole coverage will rise since the handover trigger rate of one target BS in the entire area increases. Eventually, the handover rate rises with the increasing $P_\mathrm{S^\prime}$.
It is a remarkable fact that the simulation result is lower than the theoretical analysis, which is mainly because the theoretical results are calculated by once movement for each user. While in simulation those users may have multiple trace movements inside one ERB circle, which could increase the sojourn time for the user. The other reason is that the average distance is set as the upper boundary when doing the calculation, therefore the upper boundary of theoretical curves will be higher than the simulation curves. Similar simulation results can be found in \cite{Xu2017,Teng2018}.

\begin{figure*}[!t]
    \centering
    \subfloat[The impact of $\lambda_\mathrm{S}$ when $\sigma=150$]{\includegraphics[width=0.42\textwidth]{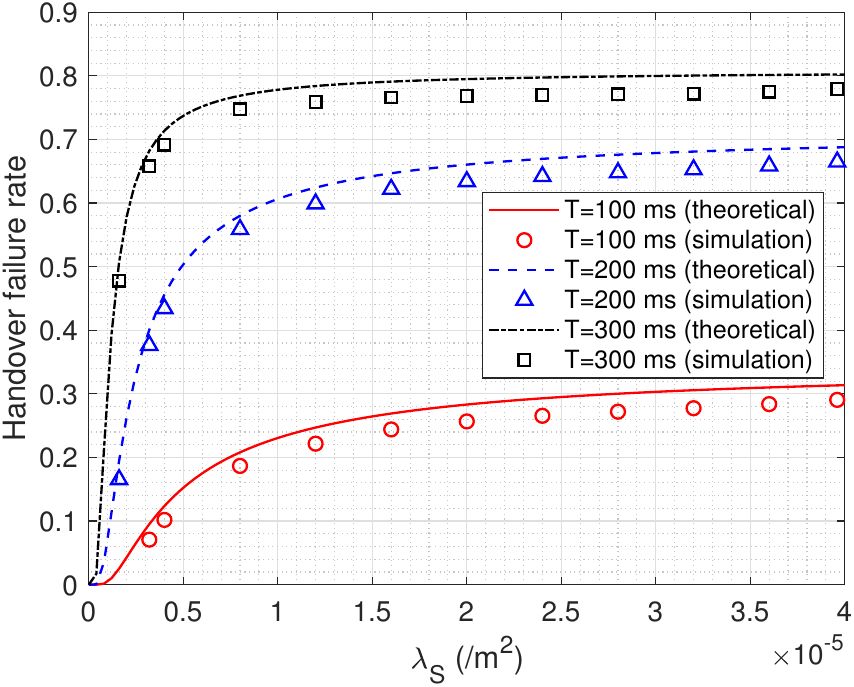}\label{fig8a}}
    \hfil
    \subfloat[The impact of $\sigma$ when $\lambda_\mathrm{S}=2\times{10}^{-5}\text{/m}^2$]{\includegraphics[width=0.42\textwidth]{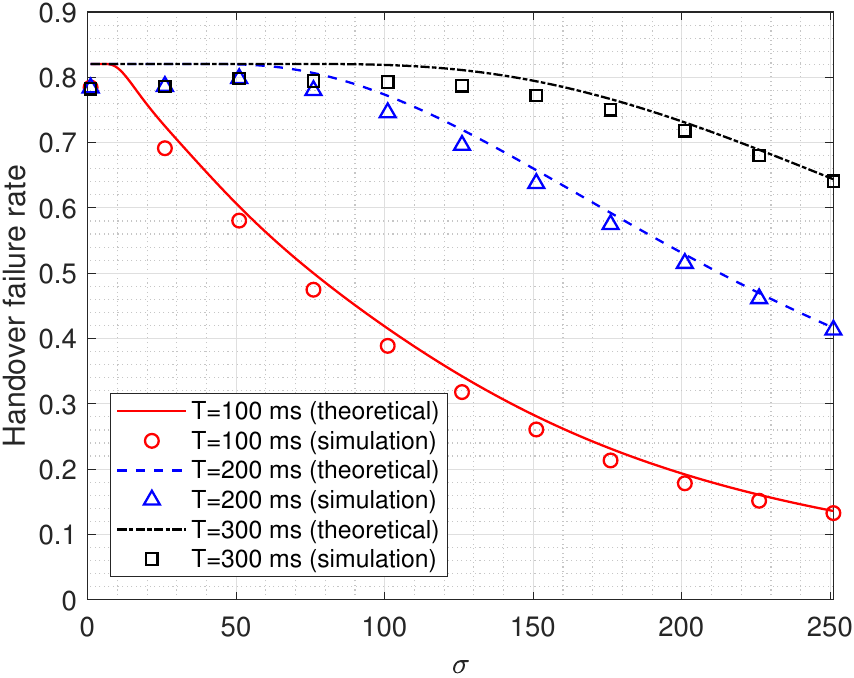}\label{fig8b}}
    \hfil
    \subfloat[The impact of $V$ when $P_\mathrm{S^\prime}=30$\,dBm]{\includegraphics[width=0.42\textwidth]{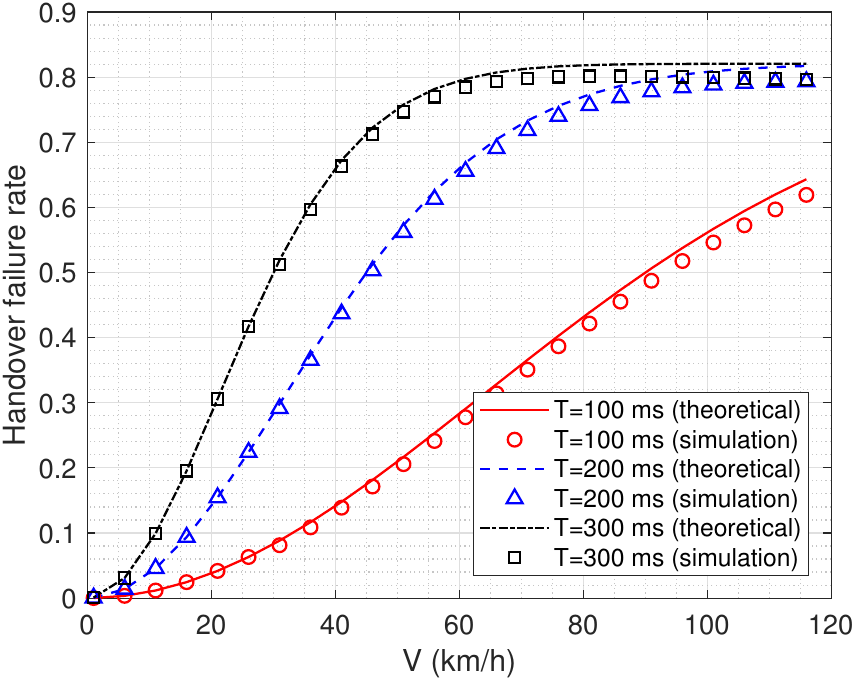}\label{fig8c}}
    \hfil
    \subfloat[The impact of $P_\mathrm{S^\prime}$ when $V=60$\,km/h]{\includegraphics[width=0.42\textwidth]{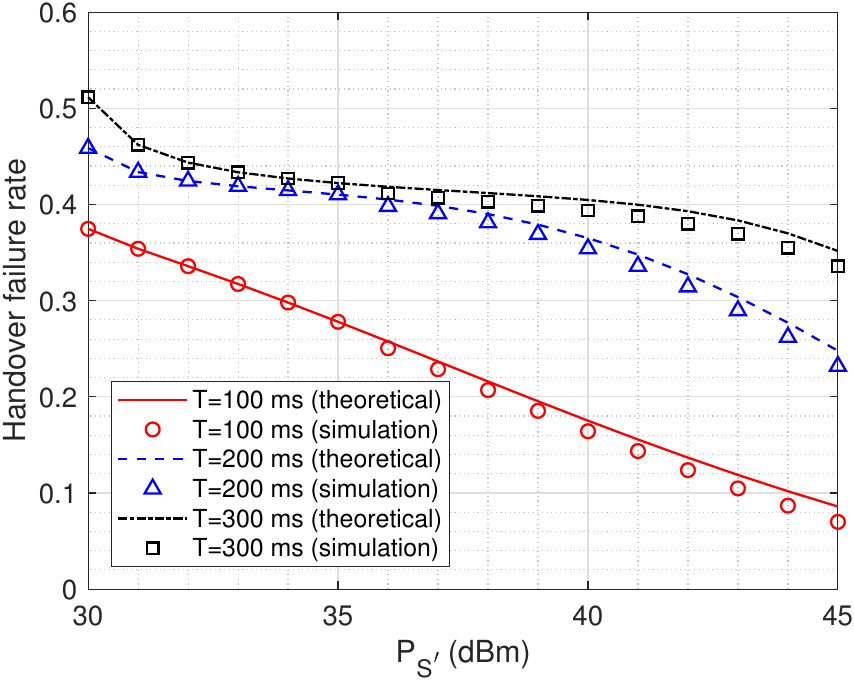}\label{fig8d}}
    \caption{The handover failure rate as a function of BS density, scattering standard variance, velocity, and transmit power at typical BS with different sojourn time thresholds.}
    \label{fig8}
\end{figure*}

Fig.~\ref{fig8} illustrates the handover failure rate as a function of $\lambda_\mathrm{S}$, $V$, $P_\mathrm{S^\prime}$, and $\sigma$ under different thresholds $T$.
As shown in Fig.~\ref{fig8a}, $H_{\mathrm{f},\ell}$ grows with the rising of $\lambda _\mathrm{S}$ and threshold $T$.
Each of the curves increases at the early stage remarkably and then slowly down the growth.
This result can be explained by the fact that the handover failure triggered rate $H_{\mathrm{f},t}$ shares the same trend with $R_{\mathrm{S^\prime S}}$, thus the handover failure rate under this condition is mainly determined by $\mathbb{P}\left( S_{\mathrm{f},\mathrm{S^\prime S}}\le T \right)$.
Since Marcum Q-function decreases remarkably at the early stage and then converges to minimum, $\mathbb{P}\left( S_{\mathrm{f},\mathrm{S^\prime S}}\le T \right)$ will follow the opposite trend with the Marcum Q-function according to \eqref{eq:S_f_ell}.
We can also see that the handover failure rate slowly reaches a stable value, which is determined by the ratio of $H_{\mathrm{f},\mathrm{t},{\mathrm{S^\prime S}}}$ and $H_{\mathrm{t},\mathrm{S^\prime S}}$.

According to Fig.~\ref{fig8b}, $H_\mathrm{f}$ decreases with the increase of $\sigma$.
The reason is that greater $\sigma$ leads to larger radius of the ERB circle, thus increasing the length of trajectory between the ERBs of handover circle and handover failure circle that a UE need to pass through when the range expansion biases are different.
Eventually, the probability that a UE touches the handover failure boundary in a unit time will decrease, which leads to a lower handover failure rate.

Fig.~\ref{fig8c} shows the trend that handover failure rate under different threshold $T$ rises with the increasing $V$. This trend can be explained by the fact that handover failure triggered rate $H_{\mathrm{f},\mathrm{t},\ell}$ rises with the increasing $V$. In the meantime, the sojourn time $S_{\mathrm{f},\ell}$ for the typical UE between handover boundary and handover failure boundary will decrease because the average distance is unchanged and the velocity is rising, which bring to the fact the $\mathbb{P} \left( S_{\mathrm{f},\ell}\le T \right)$ will also rise with the increasing $V$.

From the simulation results in Fig.~\ref{fig8d}, we can see the handover failure rate $H_{\mathrm{f},\ell}$ decrease with the increasing $P_\mathrm{S^\prime}$. 
Moreover, the influence of $P_\mathrm{S^\prime}$ on $H_{\mathrm{f},\ell}$ is different from $H_\ell$. 
During the handover failure rate analysis, the handover failure triggered rate shares the same trend with handover triggered rate. 
They will both rise with the increasing $P_\mathrm{S^\prime}$. However, the $\mathbb{P} \left( S_{\mathrm{f},\ell}\le T \right)$ decreases with the increasing $P_\mathrm{S^\prime}$, because the average distance between two ERB circles gets larger with the increasing $P_\mathrm{S^\prime}$, thus the $S_{\mathrm{f},\ell}$ will rise since the velocity of user remains unchanged. 
If $T$ remain a stable value, $\mathbb{P} \left( S_{\mathrm{f},\ell}\le T \right)$ will decrease with the increasing $P_\mathrm{S^\prime}$. 
To be noticed, the decrease of handover failure rate is nonlinear, because $\mathbb{P} \left( S_{\mathrm{f},\ell}\le T \right)$ is a Marcum Q-function and the decrease of which is nonlinear as we described in Fig.~\ref{fig7}.

\begin{figure*}[!t]
    \centering
    \subfloat[The impact of $\lambda_\mathrm{S}$ when $\sigma=150$ and $T_p=1.5$\,s.]{\includegraphics[width=0.31\textwidth]{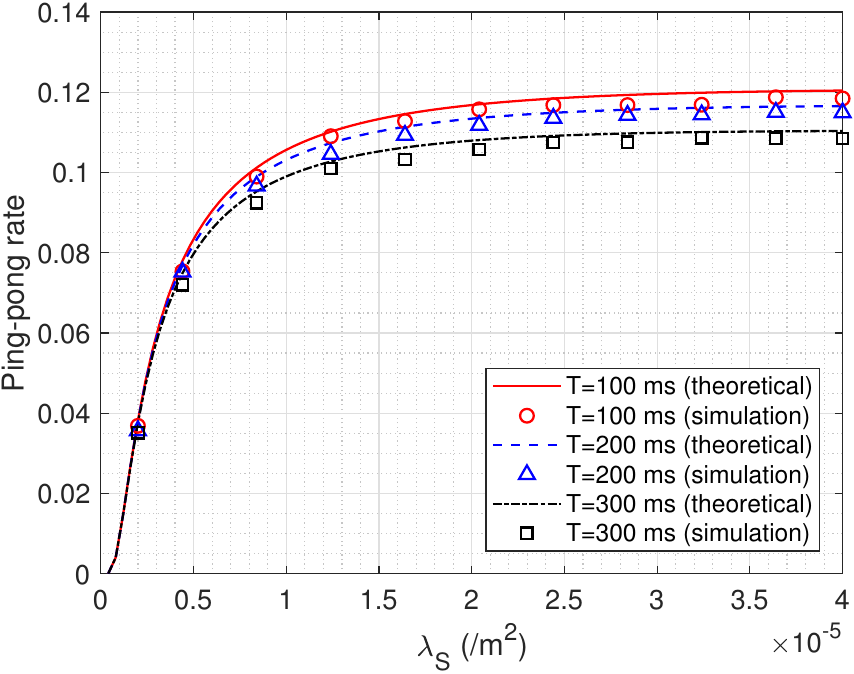}\label{fig9a}}
    \hfil
    \subfloat[The impact of $\sigma$ when $\lambda_\mathrm{S}=2\times{10}^{-5}\text{/m}^2$ and $T_p=1.5$\,s.]{\includegraphics[width=0.31\textwidth]{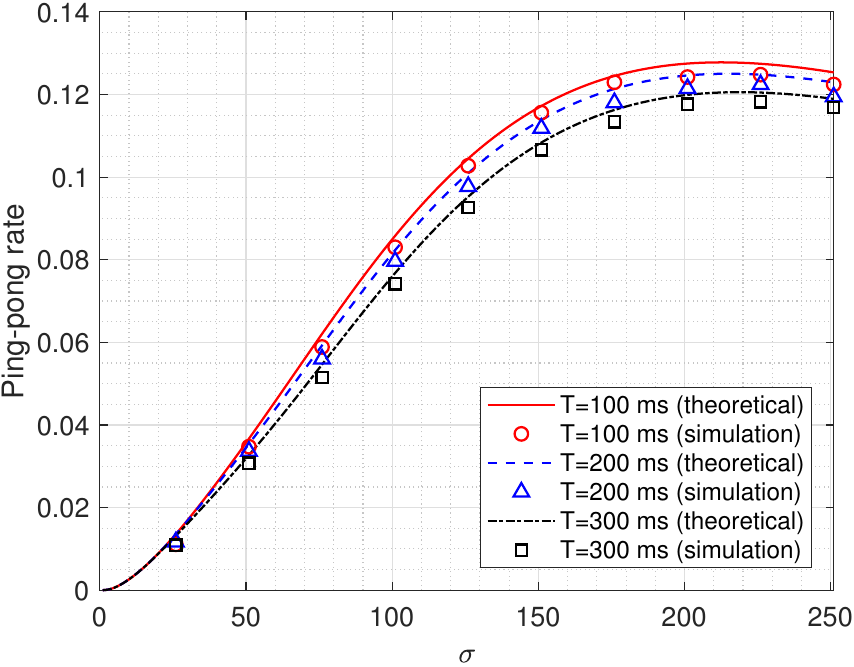}\label{fig9b}}
    \hfil
    \subfloat[The impact of $T_p$ when $\lambda_\mathrm{S}=2\times{10}^{-5}\text{/m}^2$ and $\sigma=150$.]{\includegraphics[width=0.31\textwidth]{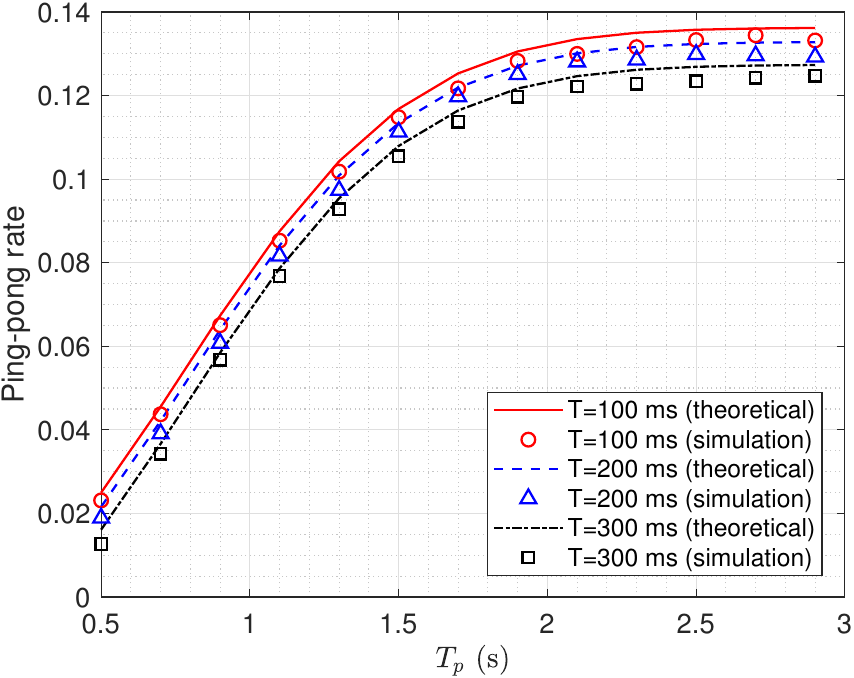}\label{fig9c}}
    \caption{The ping-pong rate as a function of BS density, scattering standard variance, and threshold $T_p$ at different sojourn time thresholds when $\mathbb{E}[N_{\mathrm{bs}}]=10$.}
    \label{fig9}
\end{figure*}

As illustrated in Fig.~\ref{fig9a}, the ping-pong rate rises and remains a much small value with the increase of $\lambda _\mathrm{S}$.
According to Fig.~\ref{fig9b}, with the increase of $\sigma$, the ping-pong rate first rises at the early stage and then slowly decreases.
It is worth noting that the ping-pong rate decreases after reaching the maximum value, which can be explained by the fact that with the increase of $\sigma$, the average distance rises, so does the corresponding radius of ERB circle.
The rising $\sigma$ will increase the average distance. In that case the ping-pong rate will decrease because when the UE moves back, it may change the moving direction before it touches the original serving BS's boundary. The longer average distance means that the UE is more likely to move into other cells.
However, the rising $\sigma$ will also increase the corresponding radius of ERB circle. For the ping-pong rate, the bigger radius indicates larger cell area, which will increase the probability that the UE touches the original boundary when it comes back.
As illustrated in Fig.~\ref{fig9c}, the ping-pong rate rises with the increase of $T_p$ and saturates to a stable value. It is obviously that larger $T_p$ will increase the time consumed by the handover decision.

\section{Conclusion}\label{sec:conclusion}

In order to derive the handover performance in HetNets under PCP scenarios, we have proposed a MRWP model to improve simulation accuracy by eliminating the DWP.
Then we have finished the analysis of the BSs locations distribution under PPP and PCP.
Based on the analytical model, we have derived the theoretical expressions of different handover metrics, including handover rate, handover failure rate, and ping-pong rate under PCP distribution scenarios, which is the first time in recent literatures. The accuracy of the proposed model and methods are validated through Monte Carlo simulations.
As illustrated in Section~\ref{sec:simulations}, those expressions are determined by deployment intensity, scattering variance of TCP, threshold of triggered time and velocity.
The results reveal that applying MRWP model in traffic hotspots can lead to better coverage performance in edge area than traditional RWP model.
Moreover, compared with the BS density, the scattering variance of small cells has more significant impact on the average distance and handover rate. The threshold $T$ should set as a small value, since smaller $T$ can significantly decreases the handover failure rate while slightly rises the handover rate and ping-pong rate. To get a balanced handover performance, $\sigma$ should be carefully selected to tradeoff the handover rate, handover failure rate, and the ping-pong rate.
Those findings in user-centric area and high density deployments networks can provide guidance for researchers to study the handover performance in ultra-dense areas. For instance the vast amounts of users' handover analysis in 5G/6G networks where multi-BSs are deployed in one small traffic hotspots.
As future work, we intend to investigate the multi-objectives optimization problems between those critical parameters to get better performance in different actual applications.

\appendices
\numberwithin{equation}{section}

\section{Proof of Lemma~\ref{lemma:lambda_optimal}}\label{appendix:lambda_optimal}
Let $q=\sqrt{x_j^{2}+y_j^{2}}$ and $F(\lambda) =\int_0^q E(r)\,\mathrm{d}r$. Recalling \eqref{eq:bound_circ_5}, $F(\lambda)$ can be formulated by
\begin{align}
    F(\lambda) =\int_0^q \left|r^{2{\alpha_{ij}}}-\lambda r^2\right|\,\mathrm{d}r.
\end{align}
Taking the derivative of $F(\lambda)$ with respect to $\lambda$ yields $F^\prime(\lambda)=-\frac{q^3}{3}+\frac{1}{3}\lambda^{\frac{3}{2\left(\alpha-1\right)}}$, $\lambda\leq r^{2\left(\alpha-1\right)}$. Then $F^{'}(\lambda)=0$ yields $\lambda=q^{2(\alpha-1)}$. When $\lambda> r^{2\left(\alpha-1\right)}$, there is no minimum value of $F(\lambda)$.

\section{Proof of Theorem~\ref{theorem:minDistancePDF}}\label{appendix:minDistancePDF}
Assume the typical BS located at origin and the hotspot center at $\left\| \mathbf{x}_0 \right\|=(x_0,y_0)\in\Phi_p$. Since the offspring SBSs $B_{\mathbf{x}_0}$ generated from $\left\| \mathbf{x}_0 \right\|$ following symmetric normal distribution with variance $\sigma^2$, the coordinate of $X\in B_{\mathbf{x}_0}$ can be determined by the PDF as follows
\begin{align}
    &f_{R_{\mathrm{S^\prime S}}|W_{\mathrm{S}}}\left( x,y\mid\left\| \mathbf{x}_0 \right\| \right)\notag
    \\
    &\qquad =\frac{1}{2\pi \sigma ^2}\exp \left\{ -\frac{\left( x-x_0 \right) ^2+\left( y-y_0 \right) ^2}{2\sigma ^2} \right\}.\label{eq:F_rw0_1}
\end{align}
The CDF of $R_{\mathrm{S^\prime S}}\triangleq\sqrt{x^2+y^2}$ is then formulated by
\begin{align}
    &F_{R_{\mathrm{S^\prime S}}|W_{\mathrm{S}}}\left( r|\left\| \mathbf{x}_0 \right\| \right)\notag
    \\
    &=\int_{\mathcal{B} \left( O,r \right)}{f_{R_{\mathrm{S^\prime S}}}\left( x,y|\left\| \mathbf{x}_0 \right\| \right) \mathrm{d}x\mathrm{d}y}\notag
    \\
    &\overset{(a)}{=}\int_0^r{\int_0^{2\pi}{rf_{R_{\mathrm{S^\prime S}}}\left( r\cos \theta ,r\sin \theta |\left\| \mathbf{x}_0 \right\| \right) \mathrm{d}\theta}\mathrm{d}r}\notag
    \\
    &=\int_0^r \frac{r}{2\pi \sigma ^2}\exp \left\{ -\frac{r^2+x_{0}^{2}+y_{0}^{2}}{2\sigma ^2} \right\}\notag
    \\
    &\qquad\times\int_0^{2\pi}{\exp \left\{ \frac{r\left( x_0\cos \theta +y_0\sin \theta \right)}{\sigma ^2} \right\} \mathrm{d}\theta}\mathrm{d}r,
\end{align}
where (a) follows by converting Cartesian into polar coordinates, i.e., $\left( x,y \right)\rightarrow \left( r\cos \theta ,r\sin \theta \right)$. Then taking the derivative of $F_{R_{\mathrm{S^\prime S}}}\left( r|\left\| \mathbf{x}_0 \right\| \right)$ with respect to $r$ yields
\begin{align}
    &f_{R_{\mathrm{S^\prime S}}|W_{\mathrm{S}}}\left( r|\left\| \mathbf{x}_0 \right\| \right) =\frac{r}{2\pi \sigma ^2}\exp \left( -\frac{r^2+x_{0}^{2}+y_{0}^{2}}{2\sigma ^2} \right)\notag
    \\
    &\qquad\qquad\times\int_0^{2\pi}\exp \left\{ \frac{r\left( x_0\cos \theta +y_0\sin \theta \right)}{\sigma ^2} \right\} \mathrm{d}\theta.
\end{align}
Let $w_0=\sqrt{x_0^2+y_0^2}$, we further have
\begin{align}
    &f_{R_{\mathrm{S^\prime S}}|W_{\mathrm{S}}}\left( r|w_0 \right)\notag
    \\
    &=\frac{r}{2\pi \sigma ^2} e^{-\frac{r^2+w_{0}^{2}}{2\sigma ^2}} \int_0^{2\pi} \exp \left\{ \frac{r\left( x_0\cos \theta +y_0\sin \theta \right)}{\sigma ^2} \right\} \mathrm{d}\theta\notag
    \\
    &\overset{(a)}{=}\frac{r}{\pi \sigma ^2}\exp \left( -\frac{r^2+w_{0}^{2}}{2\sigma ^2} \right) \int_0^{\pi}{\exp \left( \frac{rw_0\cos \theta}{\sigma ^2} \right) \mathrm{d}\theta},
\end{align}
where (a) follows from the periodicity of trigonometric function and the symmetry of cosine function. Finally, leveraging the definition of the modified Bessel function of the first kind completes the proof.

\bibliographystyle{IEEEtran}

\end{document}